\documentclass[apj]{emulateapj}

\usepackage{amsmath}
\shortauthors{LAPI ET AL.}
\shorttitle{SIZE \& KINEMATIC EVOLUTION OF MASSIVE ETGs}
\slugcomment{ACCEPTED BY ApJ}

\begin{document}

\title{The Dramatic Size and Kinematic Evolution of Massive Early-Type Galaxies}
\author{A. Lapi\altaffilmark{1,2,3}, L. Pantoni\altaffilmark{1,5}, L. Zanisi\altaffilmark{6}, J. Shi\altaffilmark{4}, C. Mancuso\altaffilmark{5},\\ M. Massardi\altaffilmark{5}, F. Shankar\altaffilmark{6}, A. Bressan\altaffilmark{1,3}, L. Danese\altaffilmark{1,2,3}}
\altaffiltext{1}{SISSA, Via Bonomea 265, 34136 Trieste, Italy}\altaffiltext{2}{INFN-Sezione di Trieste, via Valerio 2, 34127 Trieste,  Italy}\altaffiltext{3}{INAF-Osservatorio Astronomico di Trieste, via Tiepolo 11, 34131 Trieste, Italy}\altaffiltext{4}{Dept. of Astronomy, Univ. of Science and Technology of China, Hefei, 230026 Anhui, China}\altaffiltext{5}{INAF-IRA, Via P. Gobetti 101, I-40129 Bologna, Italy}
\altaffiltext{6}{Dept. of Physics and Astronomy, Univ. of Southampton, Southampton SO17 1BJ, UK}

\begin{abstract}
We aim to provide a holistic view on the typical size and kinematic evolution of massive early-type galaxies (ETGs), that encompasses their high-$z$ star-forming progenitors, their high-$z$ quiescent counterparts, and their configurations in the local Universe. Our investigation covers the main processes playing a relevant role in the cosmic evolution of ETGs. Specifically, their early fast evolution comprises: biased collapse of the low angular momentum gaseous baryons located in the inner regions of the host dark matter halo; cooling, fragmentation, and infall of the gas down to the radius set by the centrifugal barrier; further rapid compaction via clump/gas migration toward the galaxy center, where strong heavily dust-enshrouded star-formation takes place and most of the stellar mass is accumulated; ejection of substantial gas amount from the inner regions by feedback processes, which causes a dramatic puffing up of the stellar component. In the late slow evolution, passive aging of stellar populations and mass additions by dry merger events occur. We describe these processes relying on prescriptions inspired by basic physical arguments and by numerical simulations, to derive new analytical estimates of the relevant sizes, timescales, and kinematic properties for individual galaxies along their evolution. Then we obtain quantitative results as a function of galaxy mass and redshift, and compare them to recent observational constraints on half-light size $R_e$, on the ratio $v/\sigma$ between rotation velocity and velocity dispersion (for gas and stars) and on the specific angular momentum $j_\star$ of the stellar component; we find good consistency with the available multi-band data in average values and dispersion, both for local ETGs and for their $z\sim 1-2$ star-forming and quiescent progenitors. The outcomes of our analysis can provide hints to gauge sub-grid recipes implemented in simulations, to tune numerical experiments focused on specific processes, and to plan future multi-band, high-resolution observations on high-redshift star-forming and quiescent galaxies with next generation facilities.
\end{abstract}

\keywords{galaxies: evolution --- galaxies: fundamental parameters  --- galaxies: structure --- galaxies: high redshift --- galaxies: kinematics and dynamics}

\section{Introduction}\label{sec|intro}

The formation and evolution of massive early-type galaxies (ETGs) has been one of the hottest and most debated issue in the astrophysics research of the last decades.

It has been established since long times that ETGs endowed with stellar masses $M_\star\ga 3\times 10^{10}\, M_\odot$ feature homogeneous stellar populations with average ages $\ga 7-10$ Gyr, pointing toward a typical formation redshift $z\ga 1$ (e.g., Renzini 2006 and references therein). The associated star-formation efficiency $f_\star\equiv M_\star/f_b\, M_{\rm H}$, i.e., the ratio between the stellar mass to that $f_b\, M_{\rm H}\approx 0.16\, M_{\rm H}$ of the baryons originally present in the host dark matter (DM) halo is found to be substantially below unity, with values $f_\star\la 0.2$ consistently inferred from weak lensing observations (Velander et al. 2014; Hudson et al. 2015; Mandelbaum et al. 2016) and abundance matching arguments (e.g., Shankar et al. 2006; Aversa et al. 2015; Rodriguez-Puebla et al. 2015; Moster et al. 2017); on the other hand, these systems are characterized by a high stellar metallicity $Z_\star\ga Z_\odot$, with values around or even exceeding the solar one (e.g., Thomas et al. 2005; Gallazzi et al. 2014). This suggests that most of the stellar mass must be accumulated in a very intense star-formation episode, and that after quenching further gas must be hindered from infall because of heating/ejection, to avoid substantial metal dilution.

Such a picture is reinforced by the pronounced $\alpha-$enhancement observed in massive ETGs; this has to be interpreted as an iron underabundance compared to $\alpha$ elements, caused by the quenching of the star formation before Type-I$a$ supernova explosions can pollute the interstellar medium with substantial iron amounts (e.g., Thomas et al. 2005; Gallazzi et al. 2006, 2014). The implied star-formation timescales, somewhat dependent on the assumed initial mass function (IMF), boil down to a fraction of Gyr. As to the mechanism responsible for the quenching, the most likely possibility involves energy feedback from the central supermassive black hole (BH). Indeed, relic BHs are known to be hosted at the center of almost all massive ETGs, with masses $M_{\rm BH}\sim 10^{7}-10^{10}\, M_\odot$ that correlate strongly with many galaxy properties (e.g., with the stellar mass in the old stellar population, with the Sersic index of the light profile, and most fundamentally with the stellar velocity dispersion), suggesting a coevolution in the buildup of the BH and of the stellar component (see Kormendy \& Ho 2013 and references therein; Shankar et al. 2016; van den Bosch 2016).

Two relatively recent findings have shed further insight on the processes at work in the formation of massive ETGs. The first piece of news concerns the identification via deep near-IR surveys of an increasing number of quiescent massive galaxies at high redshifts $z\ga 2$ (see Ilbert et al. 2013; Duncan et al. 2014; Tomczak et al. 2014; Caputi et al. 2015; Grazian et al. 2015; Song et al. 2016; Davidzon et al. 2017; Glazebrook et al. 2017), that are found to be already in passive evolution and to feature chemical properties similar to local ETGs (e.g., Cimatti et al. 2008; van Dokkum et al. 2008), including a (super)solar metallicity and a pronounced $\alpha$-enhancement.

The second piece of news concerns the discovery of an abundant population of dusty star-forming galaxies at redshifts $z\ga 1$, which has been shown to be
responsible for the bulk of the cosmic star-formation history out to $z\la 4$
(e.g., Gruppioni et al. 2013, 2015; Rowan-Robinson et al. 2016; Bourne et al. 2017; Dunlop et al. 2017; Lapi et al. 2011, 2017a; Novak et al. 2017), and to contribute even at $z\sim 6$ (e.g., Cooray et al. 2014; Riechers et al. 2017; Strandet et al. 2017; Zavala et al. 2018; Schreiber et al. 2017a). Such an achievement has become feasible thanks to wide-area far-IR/submillimeter surveys (e.g., Lapi et al. 2011; Gruppioni et al. 2013, 2015; Weiss et al. 2013; Koprowski et al. 2014, 2016; Strandet et al. 2016), in many instances eased by gravitational lensing from foreground objects (e.g., Negrello et al. 2014, 2017; Nayyeri et al. 2016). In fact, galaxies endowed with star formation rates (SFRs) exceeding some tens $M_\odot$ yr$^{-1}$ at redshift $z\ga 2$ were largely missed by rest-frame optical/UV surveys because of heavy dust obscuration, which is difficult to correct for with standard techniques based only on UV spectral data (e.g., Bouwens et al. 2016, 2017; Mancuso et al. 2016a; Ikarashi et al. 2017; Pope et al. 2017; Simpson et al. 2017).

Follow-up optical and near-/mid-IR observations of these dusty star-forming galaxies have allowed their stellar mass content to be characterized. The vast majority features stellar masses strongly correlated with the SFR, in the way of an almost linear relationship (the so called 'main sequence') with a limited scatter around $0.25$ dex (see Rodighiero et al. 2011, 2015; Speagle et al. 2014; Salmon et al. 2015; Tasca et al. 2015; Kurczynski et al. 2016; Tomczak et al. 2016; Santini et al. 2017; Schreiber et al. 2017b). In addition, huge molecular gas reservoirs are found in these star-forming systems (Bethermin et al. 2015; Aravena et al. 2016; Scoville et al. 2014, 2016, 2017; Decarli et al. 2016; Huynh et al. 2017; Jimenez-Andrade et al. 2018) in many instances consistent with the local, integrated Schmidt-Kennicutt diagram (SFR vs. mass of molecular gas).

X-ray followup observations of dusty star-forming galaxies have revealed the growth of the central supermassive BH, before it attains a high enough mass and power to manifest as a quasar, to quench star formation and to evacuate gas and dust from the host (e.g., Alexander \& Hickox 2012 and references therein; Mullaney et al. 2012; Page et al. 2012; Johnson et al. 2013; Delvecchio et al. 2015; Rodighiero et al. 2015; Stanley et al. 2015, 2017); intriguing correlations between the nuclear power and the host stellar mass and SFR have been established. Recently, even earlier stages in the growth of the BH have been revealed by targeted X-ray observations in a gravitationally lensed, far-IR selected galaxy at $z\sim 2$ (Massardi et al. 2018).

The above findings consistently support an in situ coevolution scenario for star formation and BH accretion, envisaging these as local, time-coordinated and interlinked processes (e.g., Lapi et al. 2011, 2014, 2017b). This view is corroborated by recent studies based on the continuity equation for the stellar component of galaxies; these have demonstrated, in a closely model-independent way, that dusty star-forming galaxies constitute the progenitors of high-$z$ massive quiescent galaxies, and eventually of local massive ETGs (Mancuso et al. 2016a,b; Lapi et al. 2017b).

Here we focus on two other important aspects of ETGs and of their  quiescent and star-forming progenitors, that can offer additional clues on galaxy formation and evolution. One is related to the size of these systems, typically measured in terms of major axis (or circularized) half-light radius $R_e$. The other concerns kinematical properties, and specifically the ratio $v/\sigma$ of rotation velocity to velocity dispersion.

Local massive ETGs follow a rather tight direct relationship between half-light size $R_e$ and stellar mass $M_\star$ (see Shen et al. 2003; Cappellari et al. 2013; Cappellari 2016). Their stellar kinematics is largely dominated by random motions, with velocity ratio $(v/\sigma)_\star\la 1$ (see Cappellari et al. 2013); in particular, most of ETGs are regular rotators with $(v/\sigma)_\star\sim 0.2-1$, while a small fraction of slow rotators features values $(v/\sigma)_\star\la 0.2$. Note that, from theoretical arguments, the $(v/\sigma)_\star$ ratio is expected to depend on the intrinsic ellipticity (e.g., Binney 2005; Emsellem et al. 2007; Cappellari 2016), but marginalizing over this quantity yields a definite, albeit somewhat dispersed, relationship with stellar mass (Veale et al. 2017).

Massive quiescent progenitors of ETGs at high redshift are known to be significantly more compact, by factors $3-5$, than local ETGs at given stellar mass (Trujillo et al. 2006; Cimatti et al. 2008; van Dokkum et al. 2008; van de Sande et al. 2013; van der Wel et al. 2014; Straatman et al. 2015; Belli et al. 2014, 2017). On the other hand, kinematical studies are difficult and scarce at $z\ga 1$; the sample around $z\sim 1$ by van der Wel \& van der Marel (2008) appears to indicate an appreciably larger $(v/\sigma)_\star$ ratio with respect to local ETGs, and the trend is confirmed by the kinematical analysis of a $z\sim 2$ gravitationally lensed galaxy (Newman et al. 2015).

As to star-forming ETG progenitors, early observations in the near-IR/optical bands indicated quite large sizes, comparable to or even exceeding those of high-redshift quiescent galaxies (e.g., van Dokkum et al. 2014; Shibuya et al. 2015). However, more recent high-resolution observations in the far-IR/(sub-)mm/radio band via ground-based interferometers (including ALMA) have revealed dusty star formation to occur in a few collapsing clumps distributed over scales $\la 2$ kpc, substantially smaller than for quiescent galaxies with similar stellar mass (see Ikarashi et al. 2015; Simpson et al. 2015; Straatman et al. 2015; Barro et al. 2014; 2016a; Spilker et al. 2016; Hodge et al. 2016; Massardi et al. 2018; Tadaki et al. 2017a,b). Kinematical studies reveal the presence of a strongly baryon-dominated stellar core with high ongoing dusty SFR $\ga$ some $10^2\, M_\odot$ yr$^{-1}$, surrounded out to $\sim 15-20$ kpc by a clumpy, unstable gaseous disk in nearly Keplerian rotation (van Dokkum et al. 2015; Wisnioski et al. 2015; Burkert et al. 2016; Genzel et al. 2014, 2017; Tadaki et al. 2017a,b; Swinbank et al. 2017; Talia et al. 2018); this corresponds to a large ratio $v/\sigma\ga 3$ for the gas component, mainly determined by a substantial rotation velocity $v\ga 250$ km s$^{-1}$ and by a modest intrinsic velocity dispersion $\sigma\la 30-80$ km s$^{-1}$ related to turbulent motions (see Law et al. 2009; Genzel et al. 2011; Wisnioski et al. 2015; Turner et al. 2017; Johnson et al. 2018).

In the past literature, a variety of processes has been invoked to explain such a complex observational landscape. For example, the early growth of a gaseous clumpy disk in high redshift galaxies may be fed by cold gas streams from large-scale filaments of the cosmic web (e.g., Birnboim \& Dekel 2003; Dekel et al. 2009) or by a biased collapse of the baryons in the inner region of the halo (e.g., Romanowsky \& Fall 2012; Lapi et al. 2011, 2014, 2017; Lilly et al. 2013; Shi et al. 2017); the subsequent compaction may be triggered by violent disk instabilities (e.g., Dekel \& Burkert 2014; Bournaud 2016) or by wet mergers (e.g., Mihos \& Hernquist 1996; Hopkins et al. 2006); the quenching of the star formation and puffing up into more extended configurations may involve energy/momentum feedback from supernovae and stellar winds (e.g., White \& Frenk 1991; Cole et al. 2000; Murray et al. 2005) and from the central supermassive BH during its quasar phase (e.g., Silk \& Rees 1998; Granato et al. 2004; Fan et al. 2008, 2010; Lapi et al. 2014), or gravitational quenching (e.g., Dekel \& Birnboim 2008; Khochfar \& Ostriker 2008); finally, the late growth in size of a quiescent galaxy is thought to originate via dry merger events (e.g., Khochfar \& Silk 2006; Ciotti \& Ostriker 2007; Naab et al. 2009). We further stress that when considering the data ensemble for both high-$z$ star-forming and quiescent galaxies, a critical observational finding is that the measured sizes are significantly more scattered than for local ETGs (see Fan et al. 2008, 2010), though part of this effect can be ascribed to an observational bias for star-forming systems. Therefore a major theoretical challenge is to identify a fundamental mechanism along the ETG evolution that must account for an appreciable reduction in the size spread.

In this paper we aim to provide a physical description of the main processes responsible for the typical size and kinematics evolution of massive ETGs, including their high-redshift star-forming progenitors, their high-redshift quiescent counterparts, and their final configurations in the local Universe. Our methods will be mainly (semi-)analytic, though heavily based on detailed outcomes from state-of-the-art numerical experiments and simulations on specific processes, such as puffing up by stellar/BH feedback and dry merging. The quantitative results of our analysis will be confronted in terms of average values and dispersions with the most recent observational constraints on galaxy sizes and kinematics at different redshifts, and will provide hints to tune numerical experiments focused on specific processes.

The plan of the paper is as follows: in Sect.~\ref{sec|biased} we recall the basic notions of the biased collapse scenario for ETG formation; in Sect.~\ref{sec|size} we discuss the theoretical aspects of the size and kinematics evolution of ETG progenitors, that include gas cooling, infall and fragmentation (Sect.~\ref{sec|sizeform}), clump migration and compaction  (Sect.~\ref{sec|sizecomp}), puffing up by feedback processes (Sect.~\ref{sec|puffing}), and late-time evolution by dry mergers (Sect.~\ref{sec|merging}); quantitative results and their comparison with observations are presented in Sect.~\ref{sec|results}; finally, our findings are discussed and summarized in Sect.~\ref{sec|summary}.

Throughout this work, we adopt the standard flat cosmology
(Planck Collaboration XIII 2016) with round parameter values:
matter density $\Omega_M = 0.32$, baryon density $\Omega_b = 0.05$, Hubble
constant $H_0 = 100\, h$ km s$^{-1}$ Mpc$^{-1}$ with $h=0.67$, and mass
variance $\sigma_8 = 0.83$ on a scale of $8\, h^{-1}$ Mpc. Reported stellar masses and SFRs (or luminosities) of galaxies refer to the Chabrier (2003) IMF.

\section{Biased collapse of ETG progenitors}\label{sec|biased}

In this section we focus on basic aspects of the biased collapse scenario for the formation of ETG progenitors (see Eke et al. 2000; Fall 2002; Romanowsky \& Fall 2012; Shi et al. 2017), that will be exploited in the sequel to investigate their early evolution.

Given a halo of mass $M_{\rm H}$, we define its virial radius as
$R_{\rm H}\equiv [3\, M_{\rm H}/4\pi\,\rho_c\, \Delta_{\rm H}\, E_z]^{1/3}$, where $\rho_c\approx 2.8\times 10^{11}\, h^2\,M_\odot$ Mpc$^{-3}$ is the critical density, $\Delta_{\rm H}\simeq 18\,\pi^2+82\,[\Omega_M\,(1+z)^3/E_z-1]-39\, [\Omega_M\,(1+z)^3/E_z-1]^2$ is the nonlinear density contrast at collapse, and $E_z=\Omega_\Lambda+\Omega_M\,(1+z)^3$ is a redshift dependent factor. In the following we shall conveniently express the virial radius $R_{\rm H}$ and circular velocity $v_{c, \rm H}^2\equiv G\, M_{\rm H}/R_{\rm H}$ of the halo in terms of the stellar mass $M_\star$ enclosed in the host galaxy and of its star-formation efficiency $f_\star\equiv M_\star/f_b\,M_{\rm H}$, where $f_b\equiv \Omega_b/\Omega_{\rm M}\approx 0.16$ is the universal baryon to DM mass ratio; the outcome reads
\begin{eqnarray}\label{eq|virial}
\nonumber R_{\rm H}&\approx& 160\, f_{\star,0.2}^{-1/3}\,M_{\star,11}^{1/3}\, [E_z/E_{z=2}]^{-1/3}~~ {\rm kpc}~,\\
\\
\nonumber v_{c, \rm H}&\approx& 300\, f_{\star,0.2}^{-1/3}\,M_{\star,11}^{1/3}\, [E_z/E_{z=2}]^{1/6}~~{\rm km~s}^{-1}~,
\end{eqnarray}
where we have introduced the normalized quantities $M_{\star,11}=M_\star/10^{11}\, M_\odot$, $f_{\star,0.2}\equiv f_\star/0.2$, and we have chosen $z\approx 2$ as the fiducial formation redshift of ETG progenitors (see discussion below). The related dynamical time is
\begin{equation}
t_{\rm dyn}(R_{\rm H})\simeq {\pi\over 2}\, \sqrt{R_{\rm H}^3\over G\, M_{\rm H}}\approx 8.5\times 10^8\, [E_z/E_{z=2}]^{-1/2}~~{\rm yr}~.
\end{equation}

The star-formation efficiency $f_\star$ as a function of stellar mass $M_\star$ for central galaxies has been estimated at different redshifts basing on abundance matching techniques (see Behroozi et al. 2013; Moster et al. 2013, 2017; Aversa et 2015; Rodriguez-Puebla et al. 2015; Shi et al. 2017), and has been checked against local observations from weak lensing (e.g., Mandelbaum et al. 2016; Hudson et al. 2015; Velander et al. 2014), satellite kinematics (e.g., Wojtak \& Mamon 2013; More et al. 2011), X-ray halos around bright cluster galaxies (BCGs; Gonzalez et al. 2013; Kravtsov et al. 2014), and recently against estimates at $z\sim 1-2$ from mass profile modeling (see Burkert et al. 2016).

\begin{figure*}
\epsscale{1}\plotone{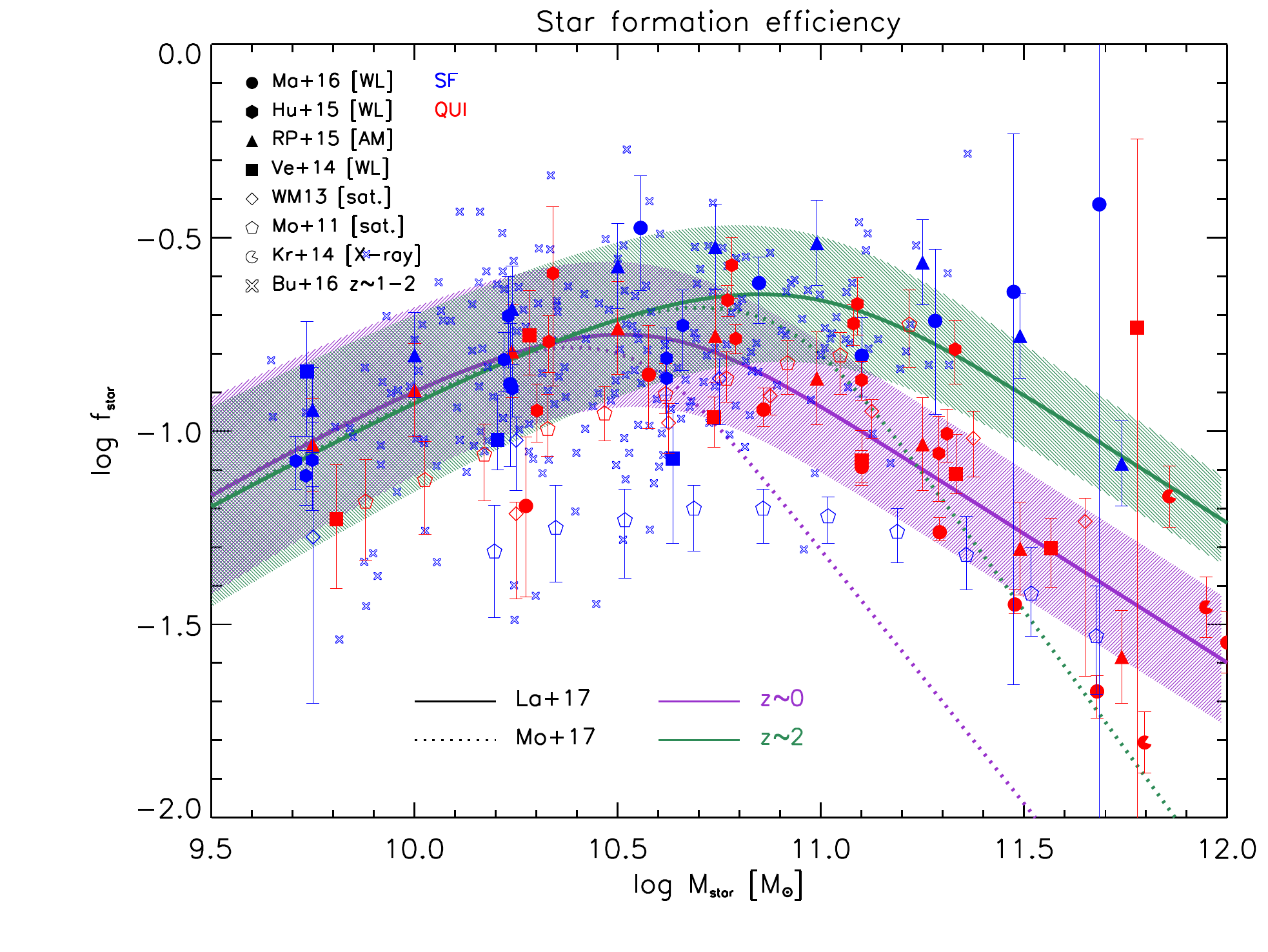}\caption{Star formation efficiency $f_{\rm star}\equiv M_\star/f_b\, M_{\rm H}$ vs. stellar mass $M_\star$. Solid lines with shaded areas illustrate the outcome by Lapi et al. (2017) via abundance matching of the halo and stellar mass functions at $z\approx 0$ (green) and $z\approx 2$ (magenta); for reference, the dotted lines refer to the results from the empirical model by Moster et al. (2017). Data points (red for quiescent and blue for star-forming galaxies) are from Mandelbaum et al. (2016; circles), Hudson et al. (2015; hexagons) and Velander et al. (2014; squares) via weak lensing, Rodriguez-Puebla et al. (2015; triangles) via subhalo abundance matching, Wojtak \& Mamon (2013; diamonds) and More et al. (2011; pentagons) via satellite kinematics, Kravtsov et al. (2014, pacmans) via X-ray observations of BCGs, and Burkert et al. (2016; crosses) via mass profile modeling at $z\sim 1-2$.}\label{fig|fstar}
\end{figure*}

The outcomes at $z\approx 0$ (magenta solid line) and $2$ (green solid line) obtained by Lapi et al. (2017) from abundance matching of the galactic halo mass function and of the stellar mass function from the continuity equation are illustrated in Fig.~\ref{fig|fstar}. For comparison, the results from an empirical model of galaxy formation by Moster et al. (2017) are also shown.
The efficiency is a nonmonotonic function of the stellar mass with maximal values $f_\star\approx 0.15-0.25$ around $M_\star\approx 10^{10}$ to a few $10^{11}\, M_\odot$, decreasing to less than $5\%$ for $M_\star\la$ a few $10^{9}\, M_\odot$ and for $M_\star\ga$ a few $10^{11}\, M_\odot$.

From the above quantities, the specific (i.e., per unit mass) angular momentum $j_{\rm H}$ of the halo is usually specified in terms of the dimensionless spin parameter $\lambda$ as
\begin{eqnarray}\label{eq|jhalo}
\nonumber  j_{\rm H} &\equiv& {\sqrt{2}\, \lambda\, R_{\rm H}\, v_{c, \rm H}}\approx 2.4\times 10^3\, \lambda_{0.035}\times\\
\\
\nonumber &\times& f_{\star,0.2}^{-2/3}\,M_{\star,11}^{2/3}\, [E_z/E_{z=2}]^{-1/6}~{\rm km~s^{-1}~kpc},
\end{eqnarray}
where $\lambda_{0.035}\equiv \lambda/0.035$. Numerical simulations (see Barnes \& Efstathiou 1987; Bullock et al. 2001; Macci\'o et al. 2007; Zjupa \& Springel 2017) have shown that $\lambda$ exhibits a log-normal distribution with average value $\langle\lambda\rangle\approx 0.035$ and dispersion $\sigma_{\log \lambda} \approx 0.25$ dex, nearly independent of mass and redshift. Moreover, the halo specific angular momentum is found to follow a radial distribution $j_{\rm H}(<r)\propto M_{\rm H}(<r)^s$ with  slope $s\approx 1$, also nearly independent on mass and redshift (e.g., Bullock et al. 2001; Shi et al. 2017).

The classic assumptions that initially the mass distribution of the baryons and of the DM mirror each other implies that $j_b(<r) = j_{\rm H}(<r)$. However, the biased collapse scenario (see Eke et al. 2000; Fall 2002; Romanowsky \& Fall 2012; Shi et al. 2017) envisages that only a fraction $f_{\rm inf}=M_{\rm inf}/f_b\, M_{\rm H}$ of the available baryons within the halo is able to cool and infall toward the central region of the galaxy where star formation takes place; under such circumstances, the specific angular momentum $j_{\rm inf}$ associated to the infalling baryons is expected to be somewhat lower than $j_{\rm H}$. In fact, one can write
\begin{eqnarray}\label{eq|jinf}
\nonumber j_{\rm inf}&=&f_{\rm inf}^s\, j_{\rm H}\approx 1.4\times 10^3\, \lambda_{0.035}\, f_{\rm inf,0.6}^s\times\\\
\\
\nonumber &\times& f_{\star,0.2}^{-2/3}\, M_{\star,11}^{2/3}\, [E_z/E_{z=2}]^{-1/6}~~{\rm km~s^{-1}~kpc}~,
\end{eqnarray}
with the normalization $f_{\rm inf,0.6}\equiv f_{\rm inf}/0.6$ discussed below. Finally, the specific angular momentum $j_\star$ retained/sampled in the local Universe by the stellar component will be a fraction of $j_{\rm inf}$.

Shi et al. (2017) have been the first to infer the infall fraction $f_{\rm inf}$ in ETGs by exploiting diverse observations on the star-formation efficiency and the chemical abundance. Basing on simple mass and metal conservation arguments, these authors found that the infall fraction can be closely estimated as
\begin{equation}\label{eq|finf}
f_{\rm inf}\simeq {y_Z\,f_\star\over Z_\star}~,
\end{equation}
in terms of the effective true metal yield of a single stellar population $y_Z$, of the star-formation efficiency $f_\star$, and of the stellar metallicity $Z_\star$. The above approximated estimate provides for ETGs outcomes to within $10\%$ accuracy relative to the exact expression derived by Shi et al. (2017; see their Sect.~3 and in particular Eqs.~11-13). However, the various quantities entering Eq.~(\ref{eq|finf}) are subject to observational and systematic uncertainties, that we now briefly discuss in turn.

As to the star-formation efficiency $f_\star$ at $z\approx 2$, we adopt the dependence on $M_\star$ and the associated scatter from Fig.~\ref{fig|fstar}. Note that the determination of the efficiency at large stellar masses $M_\star\ga 10^{12}\, M_\odot$ is rather uncertain due to difficulties in accounting for faint stellar outskirts; however, the effect should be less relevant at high redshift $z\sim 2$, since outer stellar masses are thought to be accumulated at late cosmic times $z\la 1$ via dry mergers (e.g., Rodriguez-Gomez et al. 2015, 2016; Buitrago et al. 2017).

As to the average metal yield, we adopt the fiducial value $y_Z\approx 0.069$ appropriate for a Chabrier IMF, solar metallicity and the Romano et al. (2010) stellar yield models (see also Krumholz \& Dekel 2012; Feldmann 2015; Vincenzo et al. 2016). We also allow for a systematic dispersion within the range $y_Z\sim 0.05-0.08$ that embraces values for different chemical compositions and stellar yield models (e.g., Romano et al. 2010; Nomoto et al. 2013; Vincenzo et al. 2016). It is worth noticing that for a massive galaxy formed at $z\approx 2$ with a star-formation duration of $\la 1$ Gyr, the metal yield $y_Z$ changes by less than $30\%$ from the epoch of quenching to the present time.

As to the stellar metallicity for ETGs, we adopt the average determination $Z_\star(M_\star)$ as a function of stellar mass at $z\approx 0$ and the associated scatter around $0.15$ dex by Gallazzi et al. (2014). There is clear evidence both from local massive ETGs (e.g., Choi et al. 2014; Gallazzi et al. 2006, 2014; Citro et al. 2016; Siudek et al. 2017) and from their quiescent high-redshift counterparts (e.g., Lonoce et al. 2015; Kriek et al. 2016) that, after the main burst of star formation, the metal abundance in the bulk of the stellar component stays approximately constant. Although late-time accretion of stripped stars via minor dry mergers (e.g., Rodriguez-Gomez et al. 2016; Buitrago et al. 2017) may contribute to flatten the metal gradient toward the outermost regions of local ETGs, the net effect on the average metallicity is mild (e.g., Yildrim et al. 2017; Martin-Navarro et al. 2018). Therefore, we reasonably assume that the average metallicity of present-day massive ETGs was already in place at redshift $z\sim 2$. It is also worth noticing that the determination by Gallazzi et al. (2014) is based on a combination of stellar absorption indices, that constitutes an unbiased diagnostic tool for the simultaneous derivation of ages, metallicities, and $\alpha/$Fe ratios; this dispenses with, or at least strongly alleviates, the systematic uncertainties related to measurements of metallicity from Fe abundances and hence on the occurrence rate of supernova Type-I$a$ per unit SFR (e.g., Annibali et al. 2007).

The resulting dependence of the infall fraction $f_{\rm inf}$ on the stellar mass at $z\approx 2$ is illustrated in Fig.~\ref{fig|finfout} (the stellar metallicity is plotted in the inset); it features typical values ranging from $f_{\rm inf}\approx 0.7$ to $0.6$ to $0.2$ for $M_\star$ increasing from a few $10^{10}$ to $10^{11}$ to $10^{12}\, M_\odot$, and logarithmic scatter around $\sigma_{\log f_{\rm inf}}\approx 0.25$ dex. Such behavior is indeed consistent with a scenario of biased collapse where only a fraction of the  gas initially present in the halo is processed within the central regions.

Exploiting this determination of $f_{\rm inf}$, Shi et al. (2017; cf. their Fig. 5a) have compared the predicted $j_{\rm inf}$ from Eq.~(\ref{eq|jinf}) against the locally observed $j_\star$ vs. $M_\star$ relationship for ETGs, finding a good agreement and explaining its parallel shape and lower normalization with respect to that of local spiral galaxies. Moreover, by comparing with kinematics observations (van Dokkum et al. 2015; Tadaki et al. 2017; Barro et al. 2017) of massive galaxies at $z\sim 2$, Shi et al. (2017; cf. their Fig. 5b) also confirmed an additional prediction of the biased collapse scenario, i.e., that the specific angular momentum of ETG progenitors at $z\sim 2$ is imprinted since their formation (when $\ga 70\%$ of their mass gets in place), with minor changes due to dry merging at late cosmic times (e.g., Buitrago et al. 2017; Rodriguez-Gomez et al. 2015, 2016). In conclusion, the biased collapse scenario and the associated values of $f_{\rm inf}$ are quantitatively corroborated by these two independent sets of observations.

\section{Size and kinematics of ETG progenitors}\label{sec|size}

In this Section we focus on the main processes at work in determining the size and kinematic evolution of ETG progenitors. These are schematically depicted in the cartoon of Fig.~\ref{fig|cartoon}, and comprise: biased collapse of the low angular momentum gaseous baryons located in the inner regions of the host DM halo; cooling, fragmentation, and infall of the gas down to the radius set by the centrifugal barrier; further rapid compaction via clump/gas migration toward the galaxy center, where strong and heavily dust-enshrouded star-formation activity takes place and most of the stellar mass is accumulated; ejection of substantial amount of gas from the inner regions by feedback processes and dramatic puffing up of the stellar component; passive aging of stellar populations and mass additions by dry merger events. We now turn to describe each of these processes with prescriptions inspired by basic physical arguments and by numerical simulations, and derive new analytical estimates of the relevant sizes, timescales, and kinematic properties for individual galaxies along their evolution. For definiteness, in the scaling relations of this Section we will normalize the star-formation efficiency $f_{\star}\approx 0.2$ and the infall fraction $f_{\rm inf}\approx 0.6$ to the values applying for a reference mass $M_\star\approx 10^{11}\, M_\odot$. In Sect.~\ref{sec|results} we will exploit the full mass dependence and dispersion of these quantities, to confront quantitatively our results with the available observations.

\subsection{Cooling and fragmentation}\label{sec|sizeform}

We start by computing the initial radius $R_{\rm inf}$ that encloses the infalling mass $M_{\rm inf}=f_{\rm inf}\,f_b\,M_{\rm H}$ subject to the biased collapse (see Sect.~\ref{sec|biased}). For the radial range of interest we can assume that the baryon and DM mass approximately scale with radius as\footnotemark[1]\footnotetext[1]{For a standard NFW profile (Navarro et al. 1997), the logarithmic slope of the mass distribution $M(<r)\propto r^\mu$ reads $\mu\equiv {\rm d}\log M/{\rm d}\log r = [cx/(1+cx)]^2\, [\ln (1+cx)-cx/(1+cx)]^{-1}$, in terms of the normalized radius $x\equiv r/R_{\rm H}$ and of the concentration $c$. For values $c\approx 4$ typical of massive galaxy halos virialized at $z\ga 2$ (e.g., Bullock et al. 2001; Zhao et al. 2003), the slope $\mu$ ranges from $0.8$ to $1.2$ in moving from $R_{\rm H}$ to $0.3\, R_{\rm H}$, and can be effectively approximated with unity down to $\sim 0.4-0.6\, R_{\rm H}$. For smaller radii the slope progressively approach the central value $\mu\sim 2$, which can be approximately used for $r\la 0.1\, R_{\rm H}$.} $M(<r)\propto r$, so that
\begin{eqnarray}\label{eq|Rinf}
\nonumber R_{\rm inf}&\simeq& f_{\rm inf}\, R_{\rm H}\approx 96\,f_{\rm inf,0.6}\times\\
\\
\nonumber &\times& \, f_{\star,0.2}^{-1/3}\, M_{\star,11}^{1/3}\, [E_z/E_{z=2}]^{-1/3}~{\rm kpc}~.
\end{eqnarray}
Note that the size $R_{\rm inf}$ is consistent with the scale over which both observations (see Hodge et al. 2013; Karim et al. 2013; Simpson et al. 2015; Hill et al. 2017) and high-resolution simulations (see Narayanan et al. 2015) indicate that gas, possibly segregated in multiple components, inflow toward the central regions of galaxy halos. The corresponding dynamical time reads
\begin{eqnarray}\label{eq|tdyn_Rinf}
\nonumber t_{\rm dyn}(R_{\rm inf})&\simeq& {\pi\over 2}\sqrt{R^3_{\rm inf}\over G\, M_{\rm H}(<R_{\rm inf})} = f_{\rm inf}\, t_{\rm dyn}(R_H)\, \approx\\
\\
\nonumber &\approx& 5\times 10^8\, f_{\rm inf,0.6}\,[E_z/E_{z=2}]^{-1/2}~{\rm yr}~.
\end{eqnarray}
We emphasize that a high formation redshift $z\ga 1.5$ and a low infall fraction $f_{\rm inf}\la 0.6$ enforced by the biased collapse concur to set a rather short dynamical timescale driving the subsequent evolution of ETG progenitors. For comparison, local spiral galaxies would feature a higher infall fraction $f_{\rm inf}\approx 1$ (see Shi et al. 2017) and a lower formation redshift $z\la 1$, to imply appreciably longer dynamical timescales $\ga$ a few Gyrs.

On the other hand, the radiative cooling time writes (see Sutherland \& Dopita 1993)
\begin{equation}\label{eq|tcool_Rinf}
t_{\rm cool}\simeq {2.5\times 10^{8}\over \Lambda_{-23}(T,Z)}\, {T_6\over n_{-3}\, \mathcal{C}_{10}}\,~ {\rm yr}~,
\end{equation}
where $T_6\equiv T/10^6 K$ is the temperature, $n_{-3}\equiv n/10^{-3}$ cm$^{-3}$ is the gas density, $\mathcal{C}_{10}\equiv \mathcal{C}/10$ is the clumping factor, and $\Lambda_{-23}\equiv \Lambda(T,Z)/10^{-23}$ cm$^3$ s$^{-1}$ K is the cooling function in cgs units dependent on temperature and metallicity. The normalizations have been chosen to meet the values appropriate for ETG progenitors: the infalling gas is expected to have temperatures close to the virial $T_{\rm vir}\approx 3\times 10^6\, f_{\star}^{-2/3}\, M_{\star,11}^{2/3}\, [E_z/E_{z=2}]^{1/3}$ K, and correspondingly $\Lambda_{-23}\ga 1-2$ for $Z\ga Z_\odot/10$; the gas density is expected to be of order of the average baryon density within $R_{\rm inf}$, that reads $n\approx
2\times 10^{-3}\, f_{\rm inf,0.6}^{-2}\,[E_z/E_{z=2}]\, (r/R_{\rm inf})^{-2}$ cm$^{-3}$; the clumping factor is expected to be higher than that of the IGM, which cosmological simulations (see Iliev et al. 2007; Pawlik et al. 2009; Finlator et al. 2012; Shull et al. 2012) indicate to attain values $\mathcal{C}\sim 6-20$ at $z\approx 2$. From the above it is easily understood that the cooling time $t_{\rm cool}(r)\sim 4\times 10^8\, (r/R_{\rm inf})^{-2}$ yr within $r\la R_{\rm inf}$ is comparable or shorter than the dynamical time, so that the gas can effectively cool and infall over the timescale $t_{\rm dyn}(R_{\rm inf})$. Note that such gas is rotating, being endowed with the specific angular momentum $j_{\rm inf}$ given by Eq.~(\ref{eq|jinf}).

The fraction of gas that becomes available for star formation during the infall can be addressed by looking at the fragmentation of the rotating material. Rotating discs are stable to gravitational fragmentation as far as the Toomre (1964) parameter $Q\equiv \sqrt{2}\,\Omega\,\sigma/\pi\, G \Sigma$ exceeds the critical values $0.7-1-2$ (for thick, thin, and composite discs, respectively), where $\Omega\equiv v/R\simeq j/R^2$ is the angular rotation velocity, $\sigma$ is the intrinsic velocity dispersion of the gas, generally related to turbulent motions (note that the interstellar medium is likely to become multi-phase after infall, see Braun \& Schmidt 2012), and $\Sigma\simeq M_{\rm gas}(<R)/\pi\,R^2$ is the gas surface density. The Toomre parameter can be arranged in terms of the gas mass contrast $\delta_{\rm gas}(R)\equiv M_{\rm gas}(<R)/M_{\rm tot}(<R)$, i.e. the ratio between the gas mass and the total mass (including DM), to simply read (see Dekel \& Burkert 2014)
\begin{equation}\label{eq|Q}
Q \approx {\sqrt{2}\over\delta_{\rm gas}}\,{\sigma\over v}~.
\end{equation}
The condition $Q\sim 1$ defines the stability radius $R_{\rm Q}$.

In absence of substantial fragmentation the specific angular momentum $j_{\rm inf}$ is approximately conserved (e.g., Mo et al. 1998, 2010) during contraction from the initial radius $R_{\rm inf}$ to $R_{\rm Q}$; then one finds that
\begin{eqnarray}\label{eq|RQ}
\nonumber R_{\rm Q}&\approx& {j_{\rm inf}\, Q\over \sqrt{2}\, \sigma}\,\delta_{\rm gas}(R_{\rm Q})\approx 6.3\, Q\,\sigma_{60}^{-1}\,\lambda_{0.035}\,f_{\rm inf,0.6}^s\times\\
\\
\nonumber &\times& \, f_{\star,0.2}^{-2/3}\, M_{\star,11}^{2/3}\, [E_z/E_{z=2}]^{-1/6}~{\rm kpc}~,
\end{eqnarray}
where the gas mass contrast $\delta_{\rm gas}(R_{\rm Q})\approx 0.38$ has been computed in Appendix A taking into account the effects of adiabatic contraction. In the above expression the gas intrinsic velocity dispersion $\sigma_{60}\equiv \sigma/60$ km s$^{-1}$ has been normalized to a fiducial value of $60$ km s$^{-1}$ as measured in high redshift $z\approx 2$ star-forming galaxies endowed with SFR $\ga 30\, M_{\odot}$ yr$^{-1}$ (see Law et al. 2009; Genzel et al. 2011; Wisnioski et al. 2015; Turner et al. 2017; Johnson et al. 2018).

The ratio of bulk rotation velocity to random motions after Eq.~(\ref{eq|Q}) is just
\begin{equation}\label{eq|voversigma_RQ}
\left({v\over \sigma}\right)_{\rm Q}\approx 3.7\, Q^{-1}~,
\end{equation}
and the resulting rotation velocity would approximately amount to $v_{\rm Q}\ga 200$ km s$^{-1}$ for a galaxy with stellar mass $M_\star\sim 10^{11}\, M_\odot$. Plainly, the above values are consistent with those obtained by conservation of specific angular momentum $j_{\rm inf}$ from $R_{\rm inf}$ to $R_{\rm Q}$, i.e. $v_{\rm Q}\simeq j_{\rm inf}/ k_{n}\, R_{\rm Q}$, where the constant $k_{n}\approx 1$ applies for a thick, turbulent disk with Sersic index $n\sim 1-2$ and $v/\sigma\ga 3$ (see Romanowsky \& Fall 2012; Burkert et al. 2016; Lang et al. 2017).
The corresponding dynamical time at $R_{\rm Q}$ amounts to
\begin{eqnarray}\label{eq|tdyn_RQ}
\nonumber t_{\rm dyn}(R_{\rm Q})&\simeq& {\pi\over 2}\,\sqrt{R^3_{\rm Q}\over G\, M_{\rm inf}}\approx 2.2\times 10^7\, Q^{3/2}\,\sigma_{60}^{-3/2}\,\lambda_{0.035}^{3/2}\times\\
\\
\nonumber &\times& f_{\rm inf,0.6}^{(3s-1)/2}\,f_{\star,0.2}^{-1/2}\, M_{\star,11}^{1/2}\, [E_z/E_{z=2}]^{-1/4} ~~{\rm yr}~.
\end{eqnarray}

When reaching the size $R_{\rm Q}$, the gas tends to fragment in clumps with radial velocity dispersion relative to each other of order $\sigma$. The mass of the clumps can be estimated as $M_{\rm clump}\la \pi^2\, \delta_{\rm gas}^2\, M_{\rm inf}/16\la 10^{-1}\, M_{\rm inf}$ and amounts to several percent of the disk gas mass (e.g., Bournaud et al. 2011; Dekel \& Burkert 2014), consistently with observations in high-redshift galaxies (see Elmegreen et al. 2007; Guo et al. 2018) and with the outcomes of numerical simulations (see Ceverino et al. 2010; Oklopcic et al. 2017; Mandelker et al. 2014, 2017).

In principle, gravitational torques, dynamical friction, and viscosity cooperate in order to make the gas and clumps migrating toward the inner regions (see Goldreich \& Tremaine 1980; Shlosman \& Noguchi 1993; Noguchi 1999; Immeli et al. 2004; Dekel et al. 2009; Genzel et al. 2011; also Bournaud 2016 and references therein) over a timescale
\begin{equation}\label{eq|tmigr_RQ}
t_{\rm migr}(R_{\rm Q})\simeq {2.1\, Q^2\over \delta_{\rm gas}^2(R_{\rm Q})}\, t_{\rm dyn}(R_{\rm Q})\approx 3.2\times 10^8~~{\rm yr}~.
\end{equation}
Although relevant for rotationally supported gas, this process close to $R_{\rm Q}$ is not crucial because the gravitational pull $G\, M_{\rm tot}(<R_{\rm Q})/R_{\rm Q}^2$ appreciably exceeds the centrifugal force $j_{\rm inf}^2/R_{\rm Q}^3$, or equivalently $G\, M_{\rm inf}\, R_{\rm Q}> j^2_{\rm inf}\, \delta(R_{\rm Q})$ in terms of the baryonic mass contrast $\delta(R_{\rm Q})\equiv M_{\rm inf}/M_{\rm tot}(<R_{\rm Q})\approx 0.6$ computed in Appendix A. Since rotation is not sufficient to sustain gravity, gas and clumps can continue to infall within $R_{\rm Q}$ over a dynamical time $t_{\rm dyn}(R_{\rm Q})$, while closely maintaining their initial specific angular momentum $j_{\rm inf}$ (see also Danovich et al. 2015). The infall will then be halted close to the radius where the centrifugal and gravitational forces balance (see Sect.~\ref{sec|sizecomp}).

The issue concerning the survival of clumps is extremely complex and highly debated, with both (semi-)analytical works and hydrodynamical simulations providing contrasting results, significantly dependent on sub-grid prescriptions (see Bournaud 2016 for a comprehensive review). On the one hand, it has been shown that giant clumps survive substantially intact over a few $10^8$ yr (e.g., Dekel \& Krumholz 2013; Bournaud et al. 2014; Mandelker et al. 2017); on the other hand, a number of studies suggest that an appreciable fraction of clumps can be effectively disrupted by stellar feedback over a few $10^7$ yr (e.g., Murray et al. 2010; Hopkins et al. 2012; Oklopcic et al. 2016). However, the issue is alleviated in the biased collapse scenario, because the relevant infall timescale $t_{\rm dyn}(R_{\rm Q})\sim$ a few $10^7$ yr is also quite short (see above).

During the infall, star formation proceeds in the gas (and clumps) over a timescale $t_{\rm SFR}$; observations of the correlation between star formation to gas surface density in high-redshift disks suggest values $\sim 50-100$ times longer than the dynamical time (see Elmegreen et al. 2005; Krumholz et al. 2012 and references therein), i.e.
\begin{equation}\label{eq|tsfr_RQ}
t_{\rm SFR}\simeq (50-100)\times t_{\rm dyn}(R_{\rm Q})\approx 1-2\times 10^9~~{\rm yr}~.
\end{equation}
Energy/momentum feedback via outflows from supernovae and stellar winds is expected to regulate star formation. On spatially-averaged grounds, the effects of such feedback processes are often described in terms of a mass loading factor $\epsilon_{\rm out}$, defined as the ratio between the outflow mass loss rate and the SFR (e.g., Thompson et al. 2005; Feldmann 2015); semi-analytic estimates (e.g., Lapi et al. 2014) and self-consistent hydrodynamical simulations (e.g., Hopkins et al. 2012) suggest that $\epsilon_{\rm out}\approx 1-2$ for massive galaxies with $M_\star\ga 3\times 10^{10}\, M_\odot$ of interest here. Basing on mass conservation arguments, a simple estimate of the ensuing average SFRs around $R_{\rm Q}$ reads
\begin{eqnarray}\label{eq|SFR_RQ}
\nonumber {\rm SFR}(R_{\rm Q})&\simeq& {1\over 1-\mathcal{R}+\epsilon_{\rm out}}\,{M_{\rm inf}\over t_{\rm SFR}}\la\\
\\
\nonumber &\la& 50-200\, M_\odot~{\rm yr}^{-1}~;
\end{eqnarray}
here $\mathcal{R}$ is the return fraction of gaseous material from the formed stars, taking on values $\mathcal{R}\approx 0.45$  for a Chabrier IMF (e.g., Vincenzo et al. 2016).

The above approximate analytical estimates of the SFRs, sizes $R_{\rm Q}$ and gas velocity ratios $(v/\sigma)_{\rm Q}$ are consistent with the values measured via near-IR/optical observations of $z\sim 1-2$ star-forming, massive galaxies (e.g., Genzel et al. 2014; van Dokkum et al. 2015; Barro et al. 2016a). A more quantitative comparison with data will be presented in Sect.~\ref{sec|results}.

\subsection{Compaction}\label{sec|sizecomp}

We have discussed above that, being not rotationally supported, gas and clumps can infall within $R_{\rm Q}$ over a dynamical timescale $t_{\rm dyn}(R_{\rm Q})\sim $ a few $10^7$ yr, approximately maintaining their initial specific angular momentum $j_{\rm inf}$. The process can continue down to the radius $R_{\rm rot}$ where the gravitational and centrifugal force balance
\begin{equation}
{G\, M_{\rm tot}(< R_{\rm rot})\over R_{\rm rot}^2}=\Omega^2\, R_{\rm rot}\simeq {j_{\rm inf}^2\over R_{\rm rot}^3}~.
\end{equation}
The resulting $R_{\rm rot}$ can be expressed as
\begin{eqnarray}\label{eq|Rrot}
\nonumber R_{\rm rot}&\approx& {j_{\rm inf}^2\over G\, M_{\rm inf}}\, \delta(R_{\rm rot})\approx 1.3\, \lambda_{0.035}^2\,f_{\rm inf,0.6}^{2s-1}\times\\
\\
\nonumber &\times& f_{\star,0.2}^{-1/3}\, M_{\star,11}^{1/3}\, [E_z/E_{z=2}]^{-1/3}~~{\rm kpc}~,
\end{eqnarray}
where the baryonic mass contrast is now defined as $\delta(R_{\rm rot})\equiv M_{\rm inf}/M_{\rm tot}(<R_{\rm rot})$, with typical values $\delta(R_{\rm rot})\approx 0.88$ computed in Appendix A taking into account the effects of adiabatic contraction. Eq.~(\ref{eq|Rrot}) implies an extremely high mass concentration of gas (and eventually of stars) inside $\sim 1$ kpc (see van Dokkum et al. 2014).

The kinematics at around $R_{\rm rot}$ will be dominated by rotation velocities $v_{\rm rot}\simeq j_{\rm inf}/k_{n}\, R_{\rm rot}\ga 500$  km s$^{-1}$, where the constant $k_{n}\la 2$ applies to configuration with Sersic index $n\ga 2$ and $v/\sigma\ga 3$ (see Romanowsky \& Fall 2012; Burkert et al. 2016; Lang et al. 2017). Thus the expected ratio of rotational to random motions for the gas is
\begin{eqnarray}\label{eq|voversigmagas_Rrot}
\nonumber \left({v\over \sigma}\right)_{\rm rot}&\approx& 8.9\, \sigma_{60}^{-1}\,\lambda_{0.035}^{-1}\,f_{\rm inf,0.6}^{1-s}\times\\
\\
\nonumber &\times& f_{\star,0.2}^{-1/3}\, M_{\star,11}^{1/3}\, [E_z/E_{z=2}]^{1/6}~~.
\end{eqnarray}
The dynamical time at $R_{\rm rot}$ reads
\begin{eqnarray}\label{eq|tdyn_Rrot}
\nonumber t_{\rm dyn}(R_{\rm rot}) &\simeq& {\pi\over 2}\,\sqrt{R^3_{\rm rot}\over G\, M_{\rm inf}}\approx 2\times 10^6\,\lambda_{0.035}^{3}\times\\
\\
\nonumber &\times& f_{\rm inf,0.6}^{3s-2}\, [E_z/E_{z=2}]^{-1/2} ~~{\rm yr}~.
\end{eqnarray}

The Toomre parameter at $R_{\rm rot}$ can be estimated as $Q(R_{\rm rot})\simeq \sqrt{2}/\delta_{\rm gas}(R_{\rm rot})\times (\sigma/v)_{\rm rot}$ based on Eq.~(\ref{eq|Q}); using $\delta_{\rm gas}(R_{\rm rot})\approx 0.57$ as computed in Appendix A and $(v/\sigma)_{\rm rot}\approx 9$ from Eq.~(\ref{eq|voversigmagas_Rrot}), we obtain $Q(R_{\rm rot})\approx 0.27$, a value that is pleasingly consistent with measurements in the central regions of high-$z$ star-forming galaxies (see Genzel et al. 2014). The migration time at $R_{\rm rot}$ after Eq.~(\ref{eq|tmigr_RQ}) reads
\begin{equation}\label{eq|tmigr_Rrot}
t_{\rm migr}(R_{\rm rot})\simeq {2.1\, Q^2(R_{\rm rot})\over \delta_{\rm gas}^2(R_{\rm rot})}\, t_{\rm dyn}(R_{\rm rot})\approx 9.4\times 10^5~~{\rm yr}~.
\end{equation}
Since the gas and clumps are rotationally supported at $R_{\rm rot}$, further infall can only occur by spreading out specific angular momentum via dynamical friction and gravitational torques over the above migration time. This is mirrored in the outer placement of the stellar angular momentum with respect to the stellar mass in ETGs, as noticed by Romanowsky \& Fall (2012; cf. their Fig. 2b). As $t_{\rm migr}(R_{\rm rot})$ is extremely short, the net result is a very rapid migration of the star-forming gas and clumps toward the inner regions.

Meanwhile, the star formation within $R_{\rm rot}$ occurs over a timescale
\begin{equation}\label{eq|tsfr}
t_{\rm SFR}(R_{\rm rot})\approx (50-100)\times t_{\rm dyn}(R_{\rm rot})\approx (1-2)\times 10^8~~{\rm yr}~.
\end{equation}
An estimate of the ensuing average SFR is given by
\begin{eqnarray}\label{eq|SFR}
\nonumber {\rm SFR}(R_{\rm rot})&\simeq& {1\over 1-\mathcal{R}+\epsilon_{\rm out}}\, {M_{\rm inf}\over t_{\rm SFR}(R_{\rm rot})}\la\\
\\
\nonumber &\la& 500-2000~~M_\odot\,{\rm yr}^{-1}~.
\end{eqnarray}
Thus the gas and clumps around or within $R_{\rm rot}$ are expected to feature large SFRs, rapid metal enrichment, and dust production. Note that during the early stages of this strong star-formation phase the galaxy is expected to lie above the main sequence relationship, because the stellar mass is still growing (see Mancuso et al. 2016b). Such high SFRs can partly disrupt clumps and molecular clouds (see Murray et al. 2010) and may be subject to the Eddington limit for starbursts (e.g., Andrews \& Thomson 2011; Simpson et al. 2015). All in all, we expect limited, mildly obscured SFRs in the region between $R_{\rm Q}$ and $R_{\rm rot}$, and a much stronger, obscured SFR in the innermost regions within $R_{\rm rot}$ where most of the stellar mass is accumulated; therefore the SFRs probed by UV and far-IR data are expected to be spatially disconnected (e.g., Gomez-Guijarro et al. 2018), with the UV morphology particularly knotty and irregular (e.g., Huertas-Company et al. 2015). The above approximate analytical estimates of the SFRs, sizes $R_{\rm rot}$ and velocity ratios $(v/\sigma)_{\rm rot}$ are consistent with those measured via far-IR/sub-mm and CO line observations of $z\sim 1-2$ star-forming galaxies (e.g., Barro et al. 2016a, 2017; Hodge et al. 2016; Tadaki et al. 2017; Talia et al. 2018). A more quantitative comparison with data will be performed in Sect.~\ref{sec|results}.

As $t_{\rm migr}(R_{\rm rot})\la t_{\rm dyn}(R_{\rm rot})$ violent relaxation will operate inside $R_{\rm rot}$ toward setting up a new configuration in virial equilibrium, eventually originating a bulge-like structure with Sersic index $n\ga 2$. Details of this complex process can be followed only via aimed numerical simulations (e.g., Zolotov et al. 2015; Danovich et al. 2015; Zavala et al. 2016) with apt initial conditions and space/time resolutions. The final kinematic configuration of the stars will be characterized by appreciable random motions, which for a bulge-like structure in virial equilibrium amounts to $\sigma_{\star,\rm rot}^2\simeq G\, M_\star/\beta_n\, R_{\rm rot}$ with $\beta_n\sim 4-6$ for a Sersic index $n\ga 4$. Assuming that approximately $v_{\star,\rm rot}\la v_{\rm rot}\simeq j_{\rm inf}/k_{n}\,R_{\rm rot}$ with $k_n\sim 2$ (see Romanowsky \& Fall 2012) yields a stellar velocity ratio
\begin{equation}\label{eq|voversigma_Rrot}
\left({v\over \sigma}\right)_{\star, \rm rot}\la {\sqrt{\beta_n}\over k_{n}}\, {j_{\rm inf}\over \sqrt{G\, M_\star\, R_{\rm rot}}}\approx 2\, f_{\rm inf,0.6}^{1/2}\, f_{\star,0.2}^{-1/2}~,
\end{equation}
which is substantially smaller than in the gas component; nevertheless, the system still retains appreciable rotational motions (see Barro et al. 2016b, 2017; Toft et al. 2017).

Interestingly, the compaction process described above can also create physical conditions extremely favorable to increase the gas inflow toward the innermost regions of the galaxy (from parsec to tens of parsec scale) at disposal for formation of, and rapid accretion onto a supermassive BH (e.g., Bournaud et al. 2011; Gabor \& Bournaud 2013; DeGraf et al. 2017; Rujopakarn et al. 2018). This will have important consequence for the subsequent evolution of these systems, and specifically both for the quenching of star formation and for the puffing up of the stellar distribution (see next Sect.~\ref{sec|puffing}).

We notice that at the end of the collapse, the central regions are expected to be strongly baryon-dominated. At first order, the radius $R_b$ within which baryons dominate the gravitational potential can be estimated by the equality $G\, M_{\rm inf}/R_{\rm b} \simeq G\, M_{\rm H}(<R_{\rm b})/R_{\rm b}$, which means $M_{\rm H}(<R_{\rm b})\simeq [\delta(R_{\rm b})^{-1}-1]\, M_{\rm inf}$. The result reads $R_{\rm b} \approx \sqrt{[\delta(R_{\rm b})^{-1}-1]\, 0.1\, f_{\rm inf}\, f_b}\, R_{\rm H}$, in terms of the baryonic mass contrast $\delta(R_{\rm b})\approx 0.41$ computed after taking into account adiabatic contraction (see Appendix A). Quantitatively, the baryonic-dominance radius takes on values
\begin{equation}\label{eq|R_b}
R_{\rm b}\approx 18.8\, f_{\rm inf,0.6}^{1/2}\, f_{\star,0.2}^{-1/3} \, M_{\star,11}^{1/3}\, [E_z/E_{z=2}]^{-1/3}~~{\rm kpc}~.
\end{equation}
Notice that $R_{\rm b}$ is larger than both $R_{\rm rot}$ and $R_{\rm Q}$, so that we expect a closely keplerian rotation curve out to $R_{\rm b}$, determined by the infall baryonic mass $M_{\rm inf}$. Recent observations (see van Dokkum et al. 2015; Genzel et al. 2017) and theoretical studies (see Teklu et al. 2018) are indeed revealing such a behavior.

In the above expressions for the infall radius $R_{\rm inf}$, the fragmentation radius $R_{\rm Q}$, the rotational radius $R_{\rm rot}$, and the baryon-dominance radius $R_{\rm b}$ we have normalized the star-formation efficiency $f_{\star}\approx 0.2$ and the infall fraction $f_{\rm inf}\approx 0.6$ to the values applying for a reference mass $M_\star\approx 10^{11}\, M_\odot$. When using instead the detailed dependencies on stellar mass/redshift after Figs.~\ref{fig|fstar} and \ref{fig|finfout}, we obtain the quantitative results reported in Fig.~\ref{fig|Re_Mstar_ini}; the halo size $R_{\rm H}$ computed according to Eq.~(\ref{eq|virial}) is also plotted for reference. It is seen that the mass dependence is weak, especially for $R_{\rm rot}$. For $R_{\rm Q}$ and $R_{\rm rot}$ the mild redshift evolution in the range $z\approx 1-4$ is also illustrated.

\subsection{Puffing up by feedbacks and stellar evolution}\label{sec|puffing}

An additional process contributing to alter somewhat the sizes of ETG progenitors is related to the outflow/ejection of a substantial fraction of gaseous material from the central region by feedback events (e.g., due to supernovae, stellar winds, and to the emission from the central supermassive BH during its quasar phase), that are thought to regulate or even quench star formation. As a consequence, the stellar component feels the change in the gravitational potential and relaxes to a more extended equilibrium configuration. In this process, usually refereed to as 'puffing up', the final size depends on the timescale $\tau_{\rm exp}$ of gas expulsion compared with the dynamical time $\tau_{\rm dyn}$ of the initial configuration.

For self-gravitating systems in homologous expansion, simple  arguments involving energy conservation and the virial theorem can be applied. If a fraction $f_{\rm out}$ of the infalling mass is ejected from the central star-forming regions, then the final size $R_{\rm puff}$ after puffing up is related to the initial one $R_{\rm in}$ by
\begin{eqnarray}
\nonumber {R_{\rm puff}\over R_{\rm in}} &\simeq\ \left[1-\cfrac{f_{\rm out}}{1-f_{\rm out}}\right]^{-1}~,~~~~~&\tau_{\rm exp}\ll \tau_{\rm dyn}\\
\\
\nonumber &\simeq 1+\cfrac{f_{\rm out}}{1-f_{\rm out}}~,~~~~~&\tau_{\rm exp}\ga \tau_{\rm dyn}
\end{eqnarray}
for an abrupt (Biermann \& Shapiro 1979; Hills 1980) or slow ejection (Hills 1980; Richstone \& Potter 1982), respectively; comparison of the above expressions shows that a fast ejection is more effective in increasing the size, to the point that when $f_{\rm out}\ga\ 0.5$ the system can in principle be disrupted. The corresponding velocity dispersion is expected to change from the initial $\sigma_{\rm in}$ to the final $\sigma_{\rm puff}$ value as (the quantity $\sigma^2\,R/M$ is approximately conserved for homologous expansion)
\begin{eqnarray}
\nonumber {\sigma_{\rm puff}^2\over \sigma_{\rm in}^2} &\simeq\ 1-2\,f_{\rm out}~,~~~~~&\tau_{\rm exp}\ll \tau_{\rm dyn}\\
\\
\nonumber &\simeq \left[1-f_{\rm out}\right]^2~~~~~&\tau_{\rm exp}\ga \tau_{\rm dyn}~,
\end{eqnarray}
and hence to be considerably reduced, especially in the impulsive case.
These simple results has been confirmed by numerical simulations of star clusters (e.g., Geyer \& Burkert 2001; Boily \&
Kroupa 2003; Goodwin \& Bastian 2006; Baumgardt \& Kroupa 2007; Damjanov et al. 2009), showing that the equilibrium is recovered after $20-40$ (initial) dynamical times after the ejection.

Fan et al. (2008, 2010) have been the first to suggest that such puffing up mechanism could be enforced by a massive gas outflows originated by feedback from the central supermassive BH during its quasar phase. However, in galaxies the problem is more complex due to the presence of the DM halo, and has been studied by Ragone-Figueroa \& Granato (2011) with aimed numerical experiments. They found, as expected, that the DM halo plays a stabilizing role, with two main effects. First, the size increase is appreciably reduced with respect to the analytic results above, yet still sizeable to a factor $1.5-4$ for $f_{\rm out}\ga 0.5$ (but disruption is prevented); moreover, the size increase is larger for smaller initial sizes, due to the lower contribution of DM within the central regions where the infalling baryons reside. Useful approximated formulas for the size increase and velocity dispersion decrease in presence of the DM component are given by
\begin{eqnarray}\label{eq|Rpuff}
\nonumber {R_{\rm puff}\over R_{\rm in}} &\simeq & \left(1+{\chi\,f_{\rm out}\over {1-\psi\,f_{\rm out}}}\right)~~~{\rm for}~~R_{\rm in}\approx 3~{\rm kpc}~,\\
\nonumber\\
{R_{\rm puff}\over R_{\rm in}}&\propto& \left(R_{\rm in}\over {\rm 2.7~kpc}\right)^{-\phi}~~~~~~~~{\rm for}~~f_{\rm out}\ga 0.4~,\\
\nonumber\\
\nonumber {\sigma^2_{\rm puff}\over \sigma^2_{\rm in}} &\simeq & (1-\omega\,f_{\rm out})^2~,
\end{eqnarray}
with $\chi=\psi^{-1}\approx 1.1$ and $\phi\approx 0.7$ for an impulsive ejection, $\chi=\psi\approx 0.8$ and $\phi\approx 0$ for a slow ejection, and $\omega\approx 0.7$ for both cases.

The ejected fraction $f_{\rm out}\equiv M_{\rm out}/M_{\rm inf}$ can be estimated on considering that approximately the outflown gas mass reads $M_{\rm out}\simeq M_{\rm inf}-M_\star$; the resulting
\begin{equation}\label{eq|fout}
f_{\rm out}\simeq 1-{f_{\star}\over f_{\rm inf}}\simeq 1-{Z_\star\over y_Z}~,
\end{equation}
is illustrated in Fig.~\ref{fig|finfout}, and amounts to approximately $(60\pm 10)\%$. Most of this mass loss occurs impulsively during the star-formation process due to feedbacks; in addition, a slower mass loss is related instead to stellar evolution, that restitutes a fraction $\mathcal{R}\approx 45\%$ of the material converted into stars (for a Chabrier IMF). Both these effect are taken into account in our computation.

The net outcomes are summarized in Fig.~\ref{fig|puffing}, where we show the evolution is size and velocity dispersion due to puffing up after an impulsive ejection and/or adiabatic mass loss, for different initial sizes $R_{\rm in}$ and outflowing gas fraction typical of ETG progenitors.
The size $R_{\rm puff}$ and stellar velocity ratio $(v/\sigma)_{\star, \rm puff}$ after puffing up are consistent with those measured via near-IR/optical observations of $z\sim 1-2$ quiescent galaxies (e.g., van der Wel \& van der Marel 2008; van de Sande et al. 2013; van der Wel et al. 2014; Newman et al. 2015; Hill et al. 2016; Belli et al. 2017; Glazebrook et al. 2017; Toft et al. 2017). A more detailed comparison with data will be performed in Sect.~\ref{sec|results}.

The second result by Ragone-Figueroa \& Granato (2011) concerns the timescales for equilibrium recovery, that are considerably speeded up and amount to some dynamical timescales of the region containing most of the gas.
Thus the galaxy is predicted to expand after a short time since the gas ejection, of the order of a few to tens Myr (see Eqs.~\ref{eq|tdyn_Rrot} and \ref{eq|tdyn_RQ}). This implies that the puffing up process must have been already at work in high-redshift $z\ga 2$ compact quiescent galaxies, given that the estimated age of these systems already exceeds $\ga 0.5$ Gyr (e.g., van Dokkum et al. 2009; van der Wel et al. 2014; Belli et al. 2014; Straatman et al. 2015; Toft et al. 2017; Kriek et al. 2016;  Glazebrook et al. 2017). In fact, in the past years this was an argument made against substantial puffing up by feedback processes (Damjanov et al. 2009; Ragone-Figueroa \& Granato 2011). Actually, in Section \ref{sec|results} we will see that, though compact with respect to local ETGs, $z\approx 2$ quiescent galaxies feature sizes significantly larger than for compact star-forming objects at similar redshifts, so indicating that puffing up has already affected them.
Moreover, the differential action of the puffing, which is more effective for smaller initial sizes, will turn out to be essential in reducing the large spread expected and observed in the size of compact star-forming systems, but  not seen in the rather tight size-mass relationship of local ETGs.

Note that for an efficient, impulsive puffing-up (see Ragone-Figueroa \& Granato 2011) the gas mass $M_{\rm out}\simeq f_{\rm out}\,f_{\rm inf}\,M_\star/f_\star\approx (y_Z/Z_\star-1)\,M_\star$ within $R_{\rm rot}$ must be ejected in a few dynamical times $\xi\, t_{\rm dyn}(R_{\rm rot})$ with $\xi\la 5$. The resulting mass outflow rate
\begin{equation}
\dot M_{\rm out}\approx \left({y_Z\over Z_\star}-1\right)\, {M_{\star}\over \xi\, t_{\rm dyn}(R_{\rm rot})}\sim 10^4~~M_\odot~{\rm yr}^{-1}
\end{equation}
is consistent with the values theoretically expected from feedback driven by a central supermassive BH of mass $M_{\rm BH}\ga 10^{8}\, M_\odot$ emitting close at the Eddington rate (see Granato et al. 2004; Fan et al. 2010; Lapi et al. 2006, 2014; Beckmann et al. 2017; DeGraf et al. 2017), and with the measurements for molecular and ionised winds in powerful active galactic nuclei (e.g., Chartas et al. 2009; Prochaska \& Hennawi 2009; Carniani et al. 2017; Fiore et al. 2017).

Considering the above, a specific prediction of the puffing up scenario is that on the average quasars with high SFRs ongoing in the host should feature smaller far-IR/sub-mm sizes with respect to counterparts with SFR appreciably reduced by the feedback. Such size measurements in the host galaxies of high-redshift quasars are challenging, but some data start to be collected by ALMA (see Decarli et al. 2017; Venemans et al. 2016, 2017b). In particular, the highest resolution observations (a factor $\sim 70$ better than any previous data) of a quasar at $z\sim 7.1$ with ALMA (Venemans et al. 2017b) revealed an extremely compact size $R_e\sim 1.2$ kpc of the star-forming region, as expected on the basis of our analysis (see Sect.~\ref{sec|sizeform}). On top of that, Venemans et al. (2017b) found that there is no observational evidence of significant rotational motion inside this very central regions, i.e. $v/\sigma\ll 1$. Observations of $3$ quasars at $z\ga 6.6$ set upper limits $R_e\la 4$ kpc to the size of the star-forming region, and $v/\sigma\la 1.6$ to the velocity ratio. These results support the notion that in these inner regions violent relaxation processes have been quite efficient in redistributing angular momentum outwards, as assumed in our estimate of the velocity ratio at $R_{\rm rot}$.

Finally, the puffing up can contribute to smooth out the extremely peaked stellar distribution built up during the compaction and star-formation processes, so increasing the effective Sersic index of the stellar distribution toward values $n\ga 4$, especially for the most massive galaxies. Such an effect will then be reinforced by late-time mass additions in the outskirts associated to dry merging (see below).

\subsection{Growth by dry merging}\label{sec|merging}

During the late-time evolution of ETG progenitors, the size is expected to increase because of mass additions from external dry merger events.

Following Naab et al. (2009) and Fan et al. (2010), we assume that random motions are relevant in the stellar component of quiescent ETG progenitors and set $\eta\equiv M_{\rm acc}/M_{\rm in}$ and $\epsilon\equiv \sigma^2_{\rm acc}/\sigma^2_{\rm in}$, in terms of quantities referring to the accreted and initial material. The mass after merging is therefore $M_{\rm merg} = M_{\rm in}\, (1+\eta)$. If $r\propto M^\kappa$, the virial theorem gives $\epsilon = \eta^{1-\kappa}$. Local ETGs have $\kappa\approx 0.56$ (Shen et al. 2003; Dutton et al. 2011; Lange et al. 2015) or even larger in the case of BCGs (Hyde \& Bernardi 2009); in addition, a value $\kappa\approx 0.5$ would be implied by the Faber \& Jackson (1976) relationship. From the virial theorem and the energy conservation equation, it is easily found that the fractional variations of the size and the velocity dispersion between the configurations before and after merging are
\begin{eqnarray}\label{eq|merg}
\nonumber {R_{\rm merg}\over R_{\rm in}} &=& {(1+\eta)^2\over 1+\eta^{2-\kappa}}~,\\
\\
\nonumber {\sigma_{\rm merg}^2\over \sigma_{\rm in}^2} &=& {1+\eta^{2-\kappa}\over 1+\eta}~.
\end{eqnarray}

Investigations of the fraction of close galaxy pairs and galaxies with disturbed morphologies in large catalogs (e.g., Man et al. 2016) indicate that the mass growth of massive galaxies $M_\star\ga 7\times 10^{10}\, M_\odot$ is constrained within a factor of $\sim 1.5-2$ in the redshift interval $z\sim 0.1-2.5$. Limited mass evolution $\Delta \log M_\star\approx 0.16\pm 0.04$ is also confirmed for a sample of quiescent galaxies at redshift $z\sim 1.6$ by Belli et al. (2014). Recently, Buitrago et al. (2017) have explored the assembly of the outermost regions of the most massive galaxies with $M_\star\ga 5\times 10^{10}\, M_\odot$, finding that the fraction of stellar mass stored in the outer envelopes amounts to about $30\%$ locally, an decreases to $15\%$ at $z\la 0.65$ and to $3.5\%$ at $z\sim 2$.

The analysis of the \textsl{Illustris} simulations by Rodriguez-Gomez et al. (2015, 2016) have addressed  the median fraction $f_{\rm merg}$ of ex-situ mass added by dry mergers for a given final stellar mass. The outcome is illustrated (together with the variance associated to the stochasticity in merging history) in the inset of Fig.~\ref{fig|merging}. The contribution by dry mergers is negligible for current stellar masses $M_\star\la 3\times 10^{10}\, M_\odot$ and increases appreciably for the most massive galaxies with $M_\star\ga 3\times 10^{11}\, M_\odot$, where both major and minor mergers play a relevant role; a residual small fraction of accreted mass is in the form of stars stripped from surviving galaxies, that do not originate size evolution since the required timescales are too long (see Boylan-Kolchin et al. 2008).

Basing on these numerical results, we adopt an average mass ratio $\langle\eta_{\rm M}\rangle\approx 1/4$ and $\langle\eta_{\rm m}\rangle\approx 1/10$ for major (suffix 'M') and minor (suffix 'm') mergers and compute the overall average number $\langle N_{\rm m, M}\rangle$ of major and minor mergers from $z\approx 2$ to the present time as
\begin{equation}
\langle N_{\rm m, M}\rangle = {\log [1+f_{\rm m, M}/(1-f_{\rm merg})]\over \log [1+\langle\eta_{\rm m,M}\rangle]}~;
\end{equation}
finally, we apply repeatedly Eqs.~(\ref{eq|merg}) to each merger event to obtain the global evolution in size and velocity dispersion at given final stellar mass. The outcomes are plotted in the main panel of Fig.~\ref{fig|merging}; velocity dispersion evolution is mild at all masses, while size evolution is substantial for final stellar masses $M_\star\ga 10^{11}\, M_\odot$ (see Shankar et al. 2013, 2014). These conclusions are stable against reasonable variations of the average mass ratios.

The size $R_{\rm merg}$ and stellar velocity ratio $(v/\sigma)_{\star, \rm merg}$ after dry merging are consistent with those measured via near-IR/optical observations of local ETGs (e.g., Shen et al. 2003; Cappellari et al. 2013; Cappellari 2016 and references therein). A more detailed comparison with data will be performed in Sect.~\ref{sec|results}.

\section{Results and comparison with data}\label{sec|results}

In Fig.~\ref{fig|size} and \ref{fig|sizevo} we illustrate the size vs. stellar mass relationships expected along the evolution of ETG progenitors; in particular Fig.~\ref{fig|size} offers an unified picture, while Fig.~\ref{fig|sizevo} dissects the evolution in various stages. First, we focus on star-forming progenitors. The green line refers to the fragmentation size $R_{\rm Q}$ of Eq.~(\ref{eq|RQ}), while the blue line to the rotational radius $R_{\rm rot}$ of Eq.~(\ref{eq|Rrot}). The shaded areas show the corresponding dispersions, mainly determined by that in the halo spin parameter $\lambda$; it is evident that the scatter in $R_{\rm rot}\propto \lambda^2$ is substantially larger than in $R_{\rm Q}\propto \lambda$, due to its stronger dependence. According to the discussion in Sect.~\ref{sec|size} we expect that in between the size $R_{\rm Q}$ and $R_{\rm rot}$ the typical SFRs $\la 50-200\, M_\odot$ yr$^{-1}$ are moderate and dust obscuration is mild or negligible, so that these regions can be probed by near-IR/optical observations; contrariwise,  we expect that around or within $R_{\rm rot}$ the SFRs $\la 500-2000\, M_\odot$ yr$^{-1}$ are strong and dust obscuration is heavy, so that these regions are hidden to near-IR/optical observations and can be only probed via mid/far-IR data. As an end-product of significantly larger SFRs in the central regions $\la 1$ kpc with respect to the outskirts, a very high stellar mass concentration will be originated, as indicated by observations of $z\sim 2$ massive quiescent galaxies (see van Dokkum et al. 2014).

Our expectations are consistent with the measured sizes of $z\approx 2$ star-forming galaxies (see Barro et al. 2016a, 2017; Hodge et al. 2016; Genzel et al. 2017; Tadaki et al. 2017b; Massardi et al. 2018; van Dokkum et al. 2015; Talia et al. 2018). Specifically, sizes inferred from near-IR/optical data (light blue symbols) are seen to be located in between $R_{\rm Q}$ and $R_{\rm rot}$, while sizes inferred from mid/far-IR data (dark blue symbols) lie around and within $R_{\rm rot}$. For the samples by Barro et al. (2016a) and Tadaki et al. (2017b) we have reported both the near-IR/optical size measured from HST data and the far-IR sizes from ALMA data for the very same bunch of objects, to show that the far-IR sizes are typically a factor $2-4$ smaller than the near-IR/optical ones. High-resolution, multi-band observations (e.g., Negrello et al. 2014; Massardi et al. 2018) of strongly lensed dusty star-forming galaxies have also highlighted a clear spatial segregation between the UV and far-IR emissions, with the latter being substantially more concentrated.
Note that the extremely large dispersion in the data points for star-forming galaxies is in part spuriously due to this difference between near-IR and far-IR sizes; however, even when considering data with homogenous selection, the dispersion remains substantial, in agreement with our expectation regarding the scatter on $R_{\rm Q}$ and $R_{\rm rot}$.

Note that in the literature it has been reported that the sizes of $z\sim 2$ star-forming galaxies are of the same order or even larger than that of quiescent galaxies at similar redshift (e.g., van der Wel et al. 2014; Straatman et al. 2015). However, this conclusion was based on sizes determined via near-IR/optical data, and as such it was fundamentally flawed by an observational bias. For quiescent galaxies, which are essentially dust-free, the near-IR size is a robust estimate of the radius containing most of the stellar mass; on the contrary, for strongly star-forming galaxies, which suffer of heavy dust obscuration in the inner regions, the near-IR size overestimates substantially the true radius where most of the star formation takes place and most of the stellar mass is accumulated. Taking into account high-resolution far-IR/sub-mm observations (e.g., taken with ALMA), it appears evident from the data collection in Fig.~\ref{fig|size} that the sizes of star-forming galaxies are appreciably smaller than that of quiescent galaxies.

\begin{figure*}
\epsscale{1}\plotone{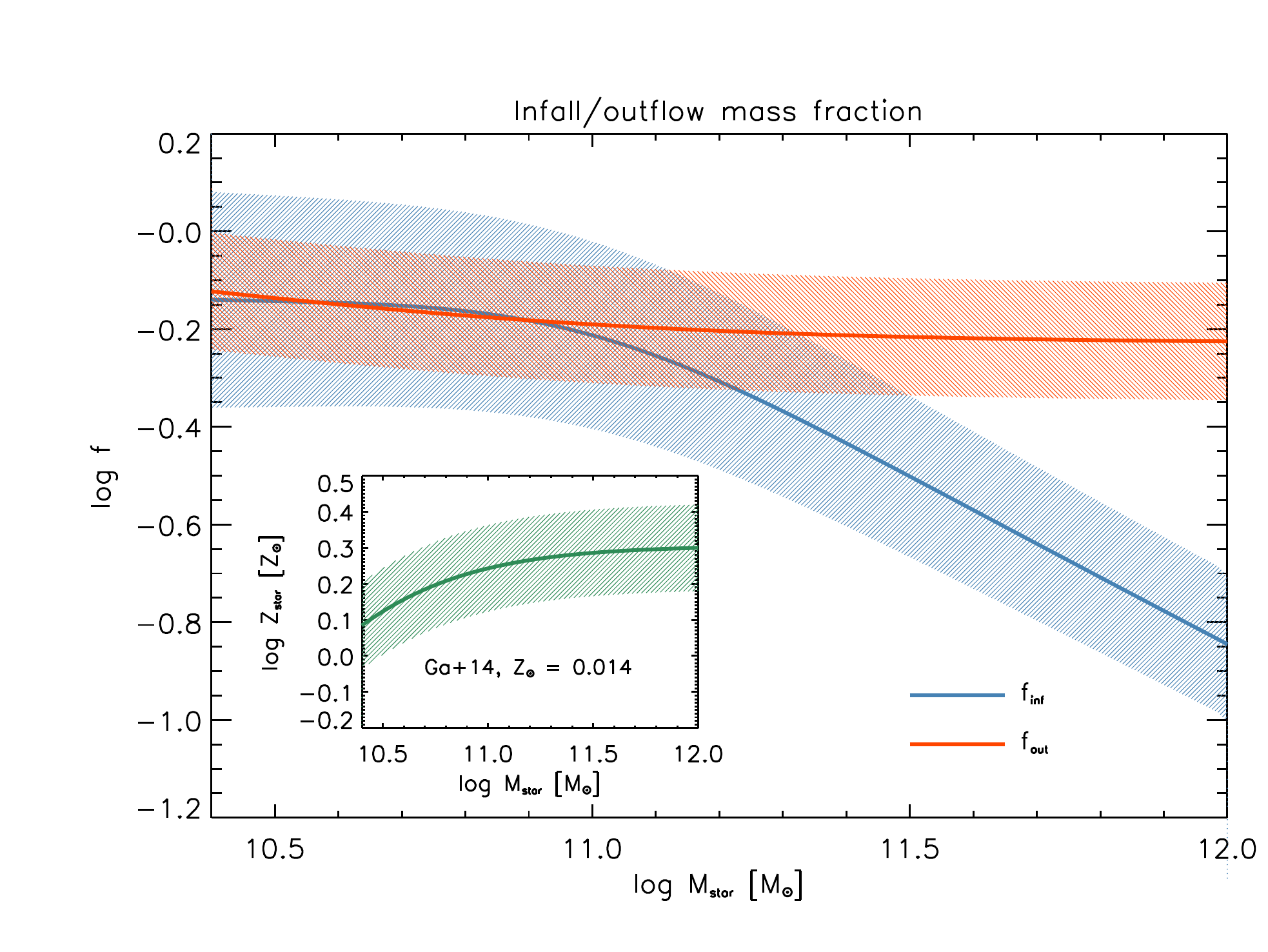}\caption{Fractions of infalling $f_{\rm inf}$ and outflowing $f_{\rm out}$ mass vs. stellar mass $M_\star$. Solid lines illustrate the average infall mass fraction $f_{\rm inf}\simeq y_Z\,f_{\rm star}/Z_\star$ (blue) and outflow mass fraction $f_{\rm out}\simeq 1-Z_\star/y_Z$ (orange), with the shaded areas showing the corresponding scatter. In the inset the solid green line illustrates the adopted average stellar metallicity $Z_\star$ vs. stellar mass $M_\star$ relationship from Gallazzi et al. (2014) for local ETGs, renormalized for a solar metallicity $Z_\odot\approx 0.014$, and the shaded area shows the associated scatter.}\label{fig|finfout}
\end{figure*}

\begin{figure*}
\epsscale{0.7}\plotone{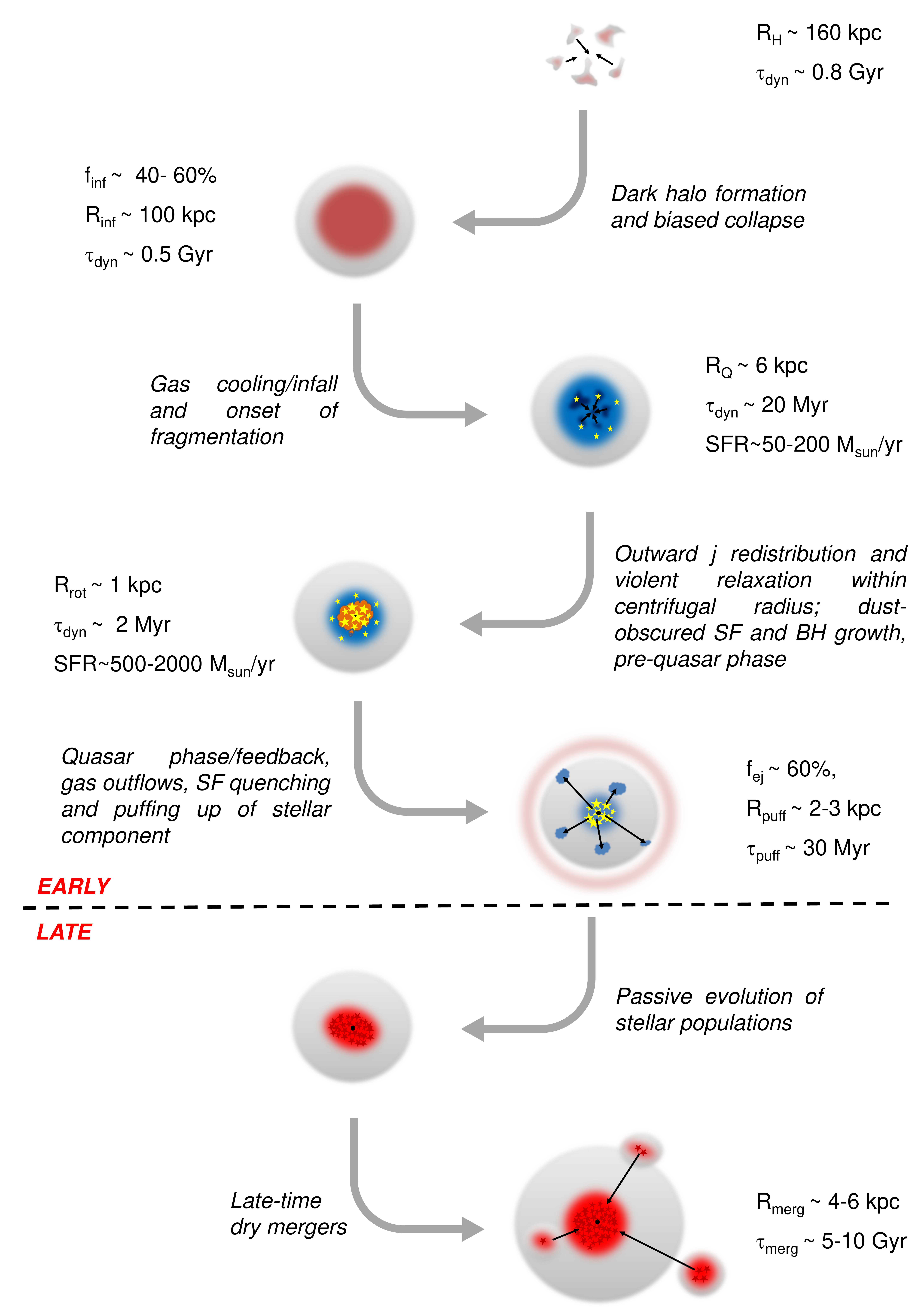}\caption{Cartoon that illustrates the main processes determining the size evolution of massive ($M_\star\sim 10^{11}\, M_\odot$) ETG progenitors, as discussed in the text. Typical sizes and timescales of the system along the evolution are also reported. The dashed horizontal line separates the early fast evolution over some $10^{8}$ yr from the late slow evolution over cosmological timescales of several Gyrs.}\label{fig|cartoon}
\end{figure*}

On the basis of Sect.~\ref{sec|puffing} we expect that after $\la$ Gyr the star formation in ETG progenitors is quenched by some feedback processes (presumably the activity of the central supermassive BH during its powerful quasar phase) and that the sudden ejection of a substantial amount of matter (see Fig.~\ref{fig|finfout}) from the central region puffs up the stellar component to a new, more extended equilibrium configuration (cf. Fig.~\ref{fig|puffing}). The resulting size $R_{\rm puff}$ illustrated in Fig.~\ref{fig|size} and \ref{fig|sizevo} as an orange solid line (the orange dashed line includes puffing up by adiabatic mass loss during passive evolution) is in agreement with the measured size of high-$z$ massive quiescent galaxies (Belli et al. 2017; Glazebrook et al. 2017; Toft et al. 2017; Hill et al. 2016; van de Sande et al. 2013; van der Wel \& van der Marel et al. 2008). Interestingly, even the sizes of local compact quiescent galaxies measured by Yildrim et al. (2017) agree well with the predicted $R_{\rm puff}$. These are galaxies stayed compact till the present, because of a lack in size evolution due to late-time dry merger events; moreover, they are known to host extremely massive BHs at their centers, that may have originated a strong puffing up at the peak time of their activity.

As discussed in Sect.~\ref{sec|puffing} and illustrated in Fig.~\ref{fig|puffing} the puffing up mechanism is more effective in galaxies with a smaller initial radius (see Eq.~\ref{eq|Rpuff}); thus the scatter associated to $R_{\rm puff}$ is found to be considerably smaller (orange shaded area) than that in $R_{\rm rot}$. This is noticeable, because the scatter in $R_{\rm rot}\propto \lambda^2$, mainly determined by that in the spin parameter $\lambda$ (see above), would have been far too large with respect to that observed in the size-mass relationships of local ETG; puffing up offers a viable mechanism to reduce the scatter in $R_{\rm rot}$ along the evolutionary sequence of ETG progenitors.

The last step in such an evolution involves the addition of mass via dry merger events, as discussed in Sect.~\ref{sec|merging}. We exploit the outcome reported in Fig.~\ref{fig|merging} for realistic mass growth histories from simulations to evolve the size of (quiescent) ETG progenitors toward the present. The resulting size $R_{\rm merg}$ is illustrated in Fig.~\ref{fig|size} and \ref{fig|sizevo} as a magenta line; the associated scatter, shown as a magenta shaded area, is somewhat increased with respect to that in $R_{\rm puff}$; this is because of the variance in the mass fraction added by dry mergers (see inset of Fig.~\ref{fig|merging}), which reflects the stochasticity in the galaxy merging histories.

The average size $R_{\rm merg}$ and its dispersion agree pretty well with the size vs. mass relationship of local ETGs as measured by the ATLAS$^{3\rm D}$ survey (dark red contours; Cappellari et al. 2013). Note that by chance the final size of ETGs are not so different from the initial fragmentation size $R_{\rm Q}$ of their progenitors, which as discussed above it is basically the size inferred via near-IR/optical observations; without the recent size measurements from far-IR/sub-mm data it would have been very difficult to envisage a self-consistent evolutionary path for ETG progenitors in the size vs. mass diagram.

Now we turn to the kinematic evolution of ETG progenitors. In Fig.~\ref{fig|vsigma} we illustrate the ratio $v/\sigma$ of the rotational velocity $v$ to the velocity dispersion $\sigma$ expected along the evolutionary history. Focusing first on the star-forming phase of ETG progenitors, we expect that the gas velocity dispersion $\sigma\approx 30-80$ km s$^{-1}$ is mainly set by turbulent motions in the interstellar medium, while the rotational velocity is associated to the angular momentum $j_{\rm inf}$ of the gas infalling in the central regions. The green line in Fig.~\ref{fig|vsigma} shows our expectation at the fragmentation size $R_{\rm Q}$, that on the basis of Eq.~(\ref{eq|voversigma_RQ}) amounts to a velocity ratio $(v/\sigma)_{\rm Q}\ga 3$, weakly increasing with stellar mass. Our predicted ratio, also taking into account the associated dispersion (green shaded area), pleasingly agrees with the current observational estimates for star-forming galaxies at $z\approx 1-2$ by Genzel et al. (2017), Di Teodoro et al. (2016), Johnson et al. 2018, Tadaki et al. (2017b), van Dokkum et al. (2015), Wisnioski et al. (2015). Note that all such dynamical observations are performed using near-IR/optical facilities (e.g., KMOS), and as such can probe reliably the $v/\sigma$ ratio for the gas only in galaxy regions where dust obscuration is not substantial, i.e., around $R_{\rm Q}$.

On the other hand, we expect that in the inner regions around $R_{\rm rot}$, where heavily dust-enshrouded star formation is ongoing and most of the stellar mass accumulates, the velocity ratio of the gas increases appreciably to values $(v/\sigma)_{\rm rot}\la 10$; this is because in the compaction from $R_{\rm Q}$ to $R_{\rm rot}$ the rotational velocity $v$ increases moderately as the specific angular momentum is nearly conserved. Preliminary data from ALMA CO observations in a couple of objects (Tadaki et al. 2017a) appear consistent with this prediction. On the other hand, the $(v/\sigma)_{\star, \rm rot}$ ratio associated to the stellar component is expected to be substantially smaller. This is because within $R_{\rm rot}$ violent relaxation processes enforce a high stellar velocity dispersion $\sigma_\star$; in the final configuration after relaxation random motions must sustain the inner gravitational potential dominated by the stellar mass. The outcome after Eq.~(\ref{eq|voversigma_Rrot}) is a ratio $(v/\sigma)_{\star, \rm rot}\sim 1-2$ nearly constant with the stellar mass. This is a specific prediction to be tested with high-resolution spectroscopic observations, albeit the strong obscuration makes the task extremely challenging while star formation is still ongoing.

The subsequent step in the evolution of ETG progenitors involves the quenching of the star formation by feedback processes and the associated puffing up. The size is increased from $R_{\rm rot}$ to $R_{\rm puff}$ by a factor a few, and as a consequence the rotational velocity is expected to decrease appreciably, while the stellar velocity dispersion is only mildly affected (see the last of Eqs.~\ref{eq|Rpuff}). The resulting velocity ratio $(v/\sigma)_{\star, \rm puff}$ for the stellar component is illustrated in Fig.~\ref{fig|vsigma} by the orange line. This is in agreement with the dynamical measurements for quiescent galaxies by Newman et al. (2015) at $z\sim 2$, by van der Wel \& van der Marel (2008) at $z\sim 1$ and by Yildrim et al. (2017) for local compact ETGs, that should actually reflect the behavior of high-redshift quiescent counterparts.

During the late time evolution of ETG progenitors we expect that the size is increased appreciably by dry merger events, especially for the most massive galaxies. From the dynamical point of view, the velocity dispersion is mildly affected (see Eq.~\ref{eq|merg} and Fig.~\ref{fig|merging}), while the rotational velocity is reduced both due to the size increase and to partial spin cancellation during encounters (e.g., Maller et al. 2002; D'Onghia \& Burkert 2004; Romanowsky \& Fall 2012); for typical mass additions of a factor $\la 1.5-2$ as occurs for massive galaxies (see Fig.~\ref{fig|merging}), the specific angular momentum loss is around $40\%$ (see Shi et al. 2017). All in all, we expect the velocity ratio $(v/\sigma)_{\star, \rm merg}$ to decrease, and especially so for the more massive galaxies that experience on average more mass additions by dry mergers. The detailed outcome based on the dry merger histories extracted from numerical simulations (see Sect.~\ref{sec|merging} and Fig.~\ref{fig|merging}) is illustrated by the magenta line in Fig.~\ref{fig|vsigma}; it agrees pleasingly with the dynamical measurements by Veale et al. (2017) from the ATLAS$^{\rm 3D}$ and the MASSIVE surveys (individual data are shown by small circles and average values by big ones) marginalized over ellipticity. In passing, it is quite interesting that the stellar velocity dispersion $\sigma_\star$ is only mildly affected both by puffing up and by late-time dry mergers; this may contribute to explain the tightness and weak evolution of the BH mass vs. $\sigma_\star$ relationship (e.g., Shankar et al. 2009; Aversa et al. 2015).

In Fig.~\ref{fig|jstar} we present the relationship between specific angular momentum and stellar mass along the evolution of ETG progenitors. Note that actually the angular momentum has been rescaled by the factor $E^{1/6}(z)$, to remove the trivial redshift evolution associated to the halo angular momentum $j_{\rm H}$, cf. Eq.~(\ref{eq|jhalo}). The green solid line and shaded area show the angular momentum $j_{\rm inf}\, E_{z=2}^{1/6}$ associated to the infalling gas in $z\sim 2$ star-forming galaxies, as expected in the biased collapse scenario for an infalling gas fraction from Eq.~(\ref{eq|finf}) and Fig.~\ref{fig|finfout}. Remarkably, our expectation for $j_{\rm inf}$ agrees well in normalization and dispersion with the measurements for $z\sim 1-2$ star-forming galaxies by Burkert et al. (2016), Tadaki et al. (2017a,b), Swinbank et al. (2017), and van Dokkum et al. (2015).

The solid magenta line with shaded area shows the stellar specific angular momentum $j_\star\, E_{z=0}^{1/6}$ expected at $z\approx 0$ after dry merger evolution, that implies both an increase in stellar mass and a $40\%$ angular momentum loss with respect to the initial $j_{\rm inf}$ by partial spin cancellation during encounters (see Sect.~\ref{sec|merging}); for comparison, the angular momentum $j_\star\, E_{z=0}^{1/6}$ at $z\approx 0$ with no momentum loss (but still including the increase in stellar mass) is shown as a dashed magenta line. The outcome for $j_\star$ is in reasonable agrement (given the large scatter) with the data for quiescent galaxies at $z\sim 1-2$ by Toft et al. (2017), Newman et al. (2015), and van der Wel \& van der Marel (2008), and for local ETGs by Romanowsky \& Fall (2012).

\section{Discussion and conclusions}\label{sec|summary}

In this paper we have provided a holistic view on the typical size and kinematic evolution of massive ETGs, that encompasses their high-$z$ star-forming progenitors, their high-$z$ quiescent counterparts, and their configurations in the local Universe.

Our investigation covers the main processes playing a relevant role in the cosmic evolution of ETGs. Specifically, their early fast evolution comprises: biased collapse of the low angular momentum gaseous baryons located in the inner regions of the host DM halo; cooling, fragmentation, and infall of the gas down to the radius set by the centrifugal barrier; further rapid compaction via clump/gas migration toward the galaxy center, where strong heavily dust-enshrouded star-formation activity takes place and most of the stellar mass is accumulated; ejection of substantial amount of gas from the inner regions by feedback processes and dramatic puffing up of the stellar distribution. In the late slow evolution, passive aging of stellar populations and mass additions by dry merger events occur.

We have described these processes relying on prescriptions inspired by basic physical arguments and by numerical simulations, to derive new analytical estimates of the relevant sizes, timescales, and kinematic properties for individual galaxies along their evolution. Then we have obtained quantitative results as a function of galaxy mass and redshift and have compared them to recent observational constraints on half-light size $R_e$, on the ratio $v/\sigma$ between rotation velocity and velocity dispersion (for gas and stars) and on the specific angular momentum $j_\star$ of the stellar component; we have found an overall good consistency with the available multi-band data in average values and dispersion both for local ETGs and for their $z\sim 1-2$ star-forming and quiescent progenitors.

Our main conclusions are the following.

\begin{figure*}
\epsscale{1.}\plotone{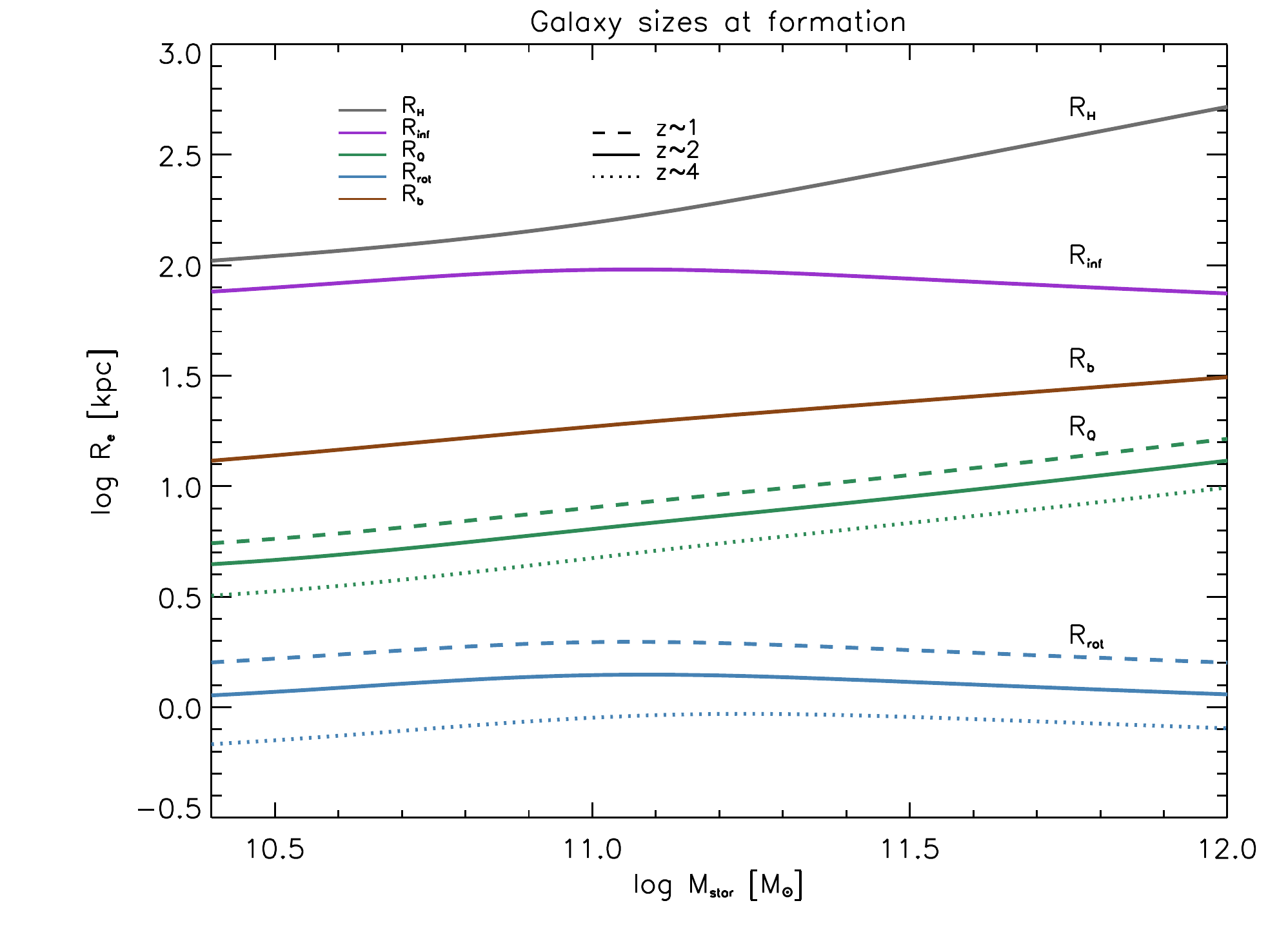}\caption{Relevant sizes $R_e$ vs. stellar mass $M_\star$ for ETG progenitors at $z\approx 2$. Gray lines refer to
the dark halo size $R_{\rm H}$ of Eq.~(\ref{eq|virial}), purples lines to the infall size $R_{\rm inf}$ of Eq.~(\ref{eq|Rinf}), green line to the fragmentation size $R_{\rm Q}$ of Eq.~(\ref{eq|RQ}), blue line to the centrifugal size $R_{\rm rot}$ of Eq.~(\ref{eq|Rrot}), and brown line to the baryonic-dominance size $R_{\rm b}$ of Eq.~(\ref{eq|R_b}). For $R_{\rm Q}$ and $R_{\rm rot}$, solid lines show the sizes expected at redshift $z\approx 2$, dashed lines at $z\approx 1$ and dotted lines at $z\approx 4$.}\label{fig|Re_Mstar_ini}
\end{figure*}

\begin{figure*}
\epsscale{1}\plotone{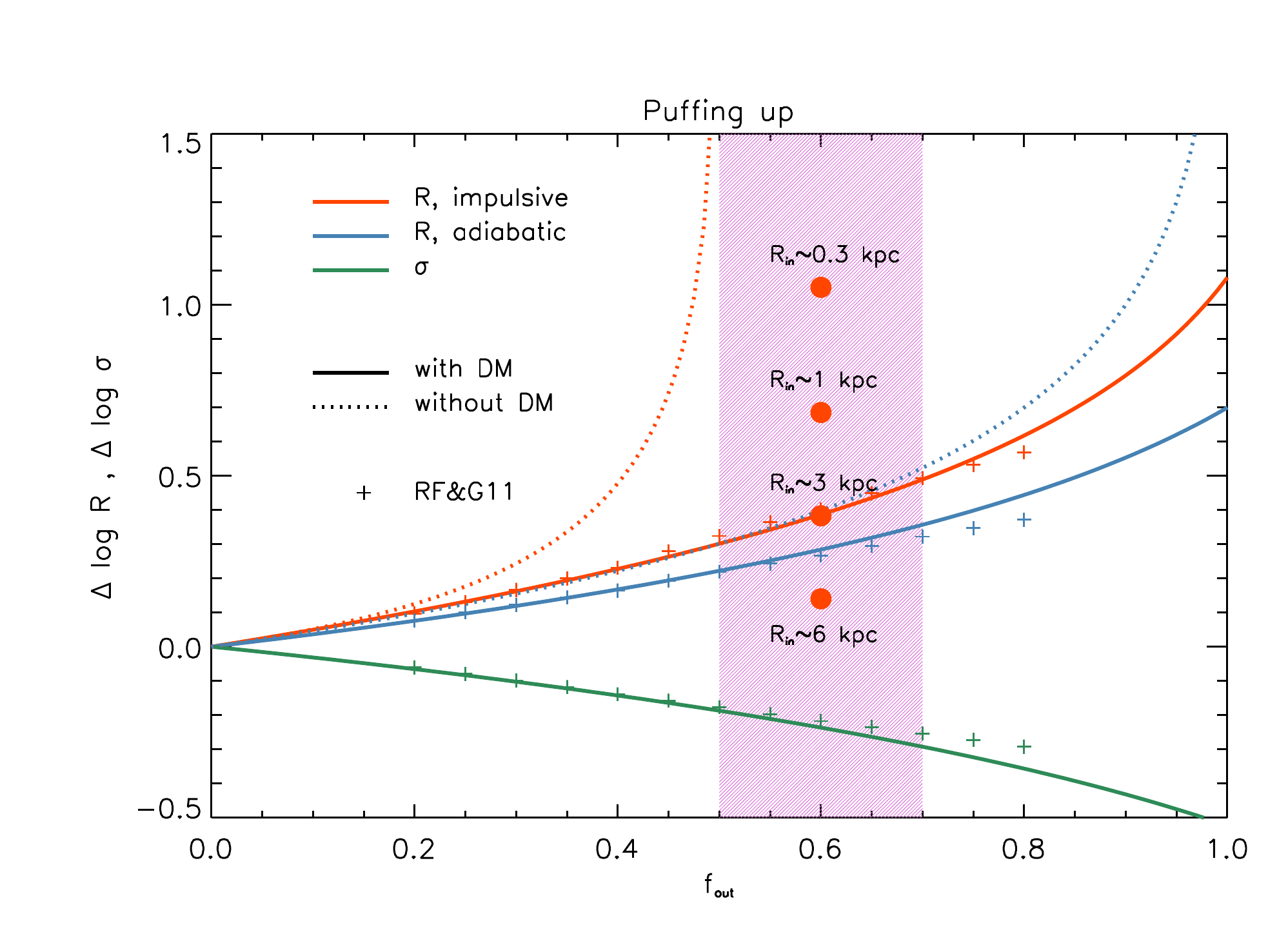}\caption{Size and velocity dispersion evolution due to puffing up, in terms of logarithmic changes $\Delta\log R$ and $\Delta\log\sigma$ as a function of the fraction of outflown gas mass $f_{\rm out}$. Orange lines and symbols refer to size evolution due to an impulsive ejection, blue lines and symbols to size evolution due to an adiabatic mass loss, and green lines and symbols to velocity dispersion evolution (the impulsive and adiabatic cases practically coincide). Small crosses are the outcomes from the numerical experiments by Ragone-Figueroa \& Granato (2011) in the presence of DM and for an initial size $R_{\rm in}\approx 3$ kpc, solid lines illustrates our analytic rendition, and dotted lines show for reference the classic result for self-gravitating systems in absence of DM; filled circles illustrate how the size increase for $f_{\rm out}\approx 0.6$ is affected when starting from different initial radii $R_{\rm in}\approx 6$, $3$, $1$, and $0.3$ kpc. The shaded magenta area reports the range of outflown mass fraction expected in ETG progenitors according to Eq.~(\ref{eq|fout}) and Fig.~\ref{fig|finfout}.}\label{fig|puffing}
\end{figure*}

\begin{figure*}
\epsscale{1}\plotone{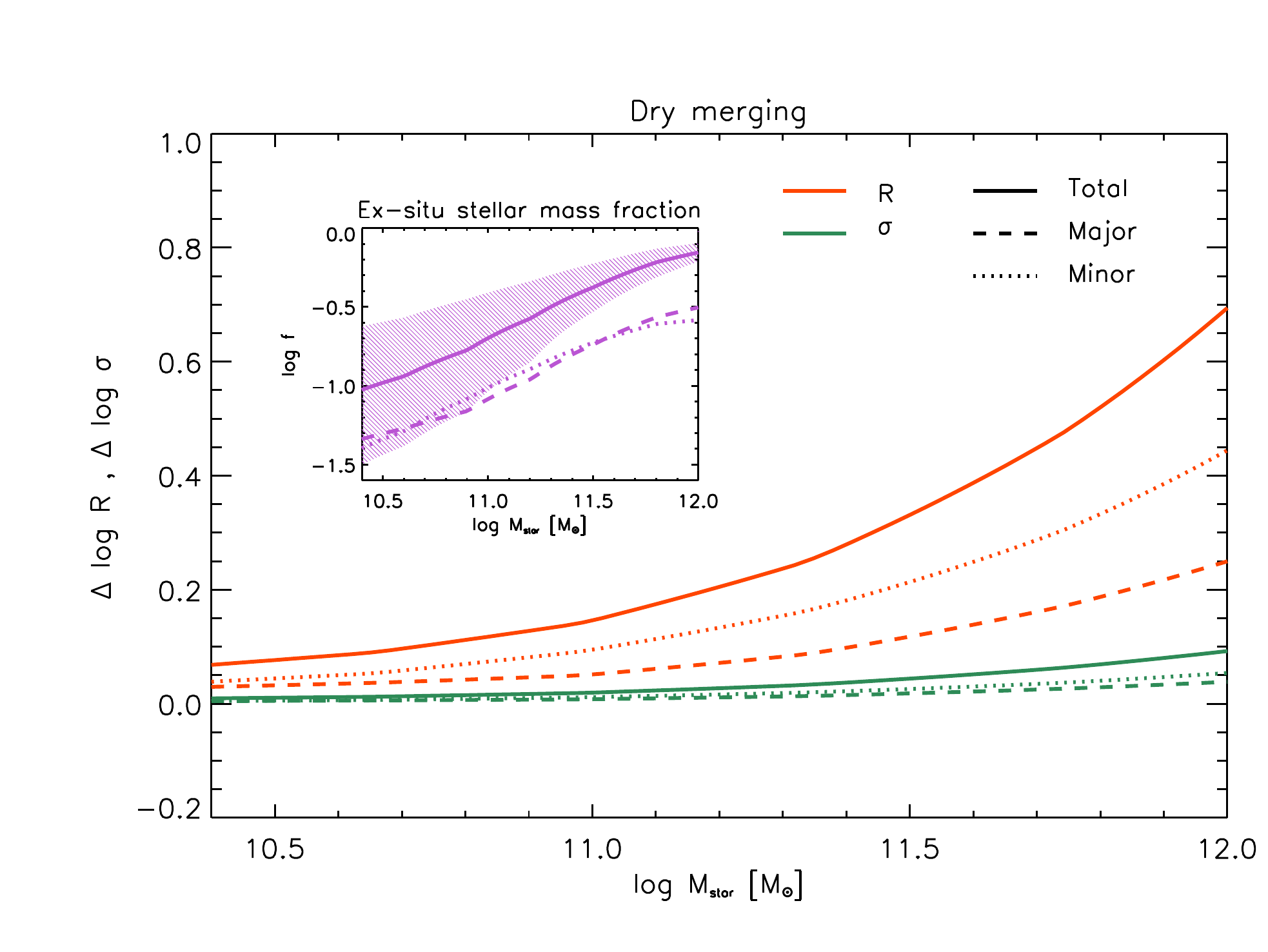}\caption{Size and velocity dispersion evolution due to dry mergers, in terms of logarithmic changes $\Delta\log R$ and $\Delta\log\sigma$ as a function of the final stellar mass $M_\star$. Orange lines and symbols refer to size evolution, and green lines to velocity dispersion evolution. Solid lines illustrates the overall evolution, dashed lines the evolution due to major dry mergers, and dotted lines the evolution due to minor mergers. These outcomes are based on the ex-situ stellar mass fraction provided by the analysis of the \textsl{Illustris} simulations by Rodriguez-Gomez et al. (2016), which is illustrated in the inset. Solid, dashed and dot-dashed linestyles as above for the overall, major and minor merger fractions, respectively; the shaded area shows the variance associated to the stochasticity of the merging histories.}\label{fig|merging}
\end{figure*}

\begin{figure*}
\epsscale{1}\plotone{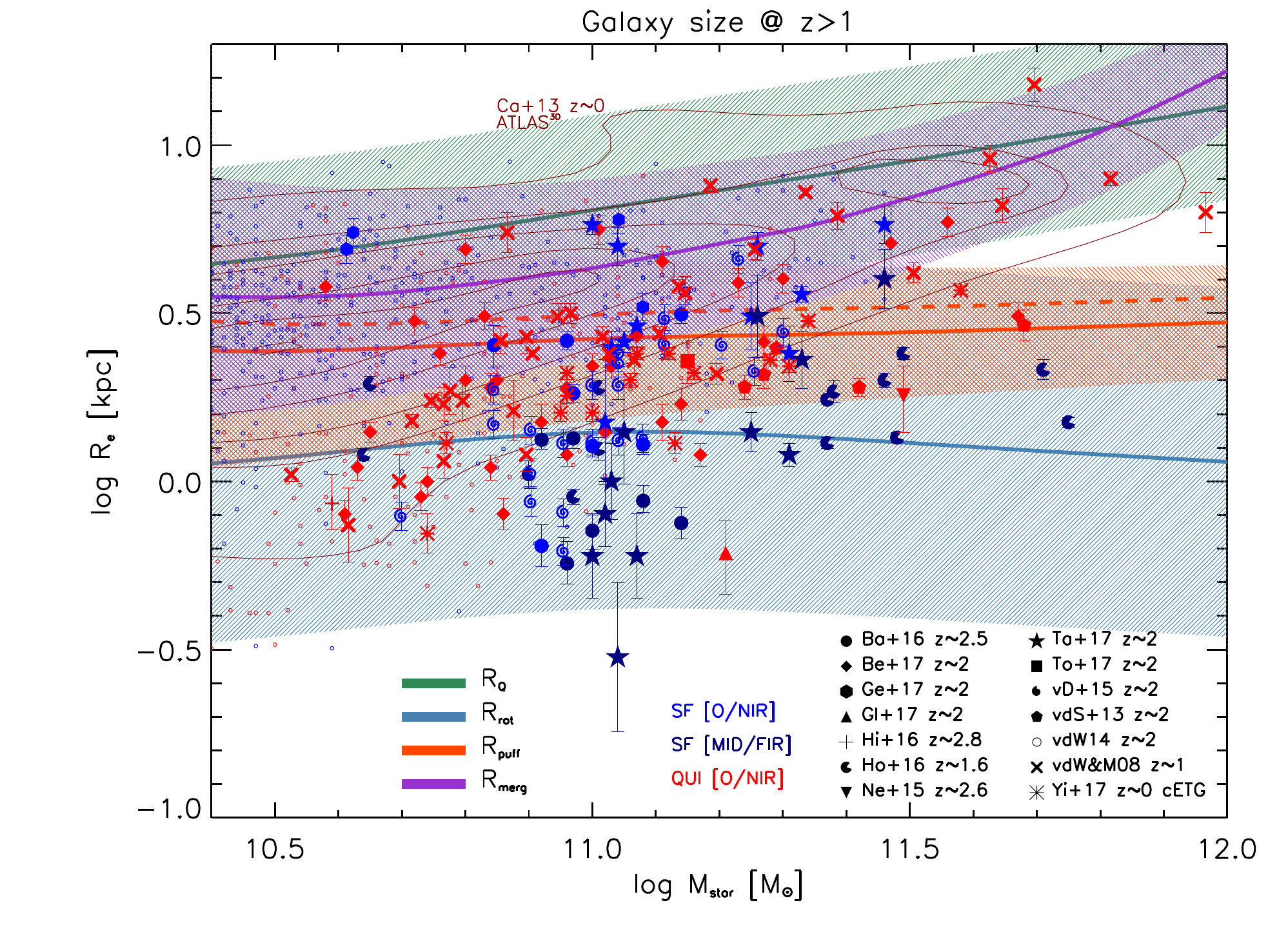}\caption{Size $R_e$ vs. stellar mass $M_\star$ relationship. The colored lines with shaded areas illustrate the size-mass relationship and the associated scatter expected along the evolution of ETG progenitors: green line refers to the fragmentation size $R_{\rm Q}$ of Eq.~(\ref{eq|RQ}), blue line refers to the centrifugal size $R_{\rm rot}$ of Eq.~(\ref{eq|Rrot}), orange line to the size $R_{\rm puff}$ of Eq.~(\ref{eq|Rpuff}) after puffing up (solid for impulsive puffing due to feedbacks and dashed for adiabatic puffing due to stellar evolution), and magenta line to the final size $R_{\rm merg}$ after dry merging. Data are from Barro et al. (2016a; circles) at $z\sim 2.5$, Belli et al. (2017; diamonds) at $z\sim 2$, Genzel et al. (2017; hexagons) at $z\sim 2$, Glazebrook et al. (2017; triangles) at $z\sim 2$, Hill et al. (2016; plus sign) at $z\sim 2.8$, Hodge et al. (2016; pacmans) at $z\sim 1.6$, Newman et al. (2015; reverse triangles) at $z\sim 2.6$, Tadaki et al. (2017b; stars) at $z\sim 2$, Toft et al. (2017; squares) at $z\sim 2$, van Dokkum et al. (2015; spirals) at $z\sim 2$, van de Sande et al. (2013; pentagons) at $z\sim 2$, van der Wel \& van der Marel (2008; crosses) at $z\sim 1$, van der Wel et al. (2014; dots) at $z\sim 2$, Yildrim et al. (2017; asterisks) at $z\sim 0$ for compact ETGs. Blue points refer to star-forming galaxies (light blue for sizes inferred from optical/near-IR observations, deep blue for sizes inferred from mid-/far-IR observations), and red points refer to quiescent galaxies. The dark red contours report the size distributions of local ETGs from the ATLAS$^{3\,\rm D}$ survey by Cappellari et al. (2013).}\label{fig|size}
\end{figure*}

\begin{figure*}
\epsscale{0.55}\plotone{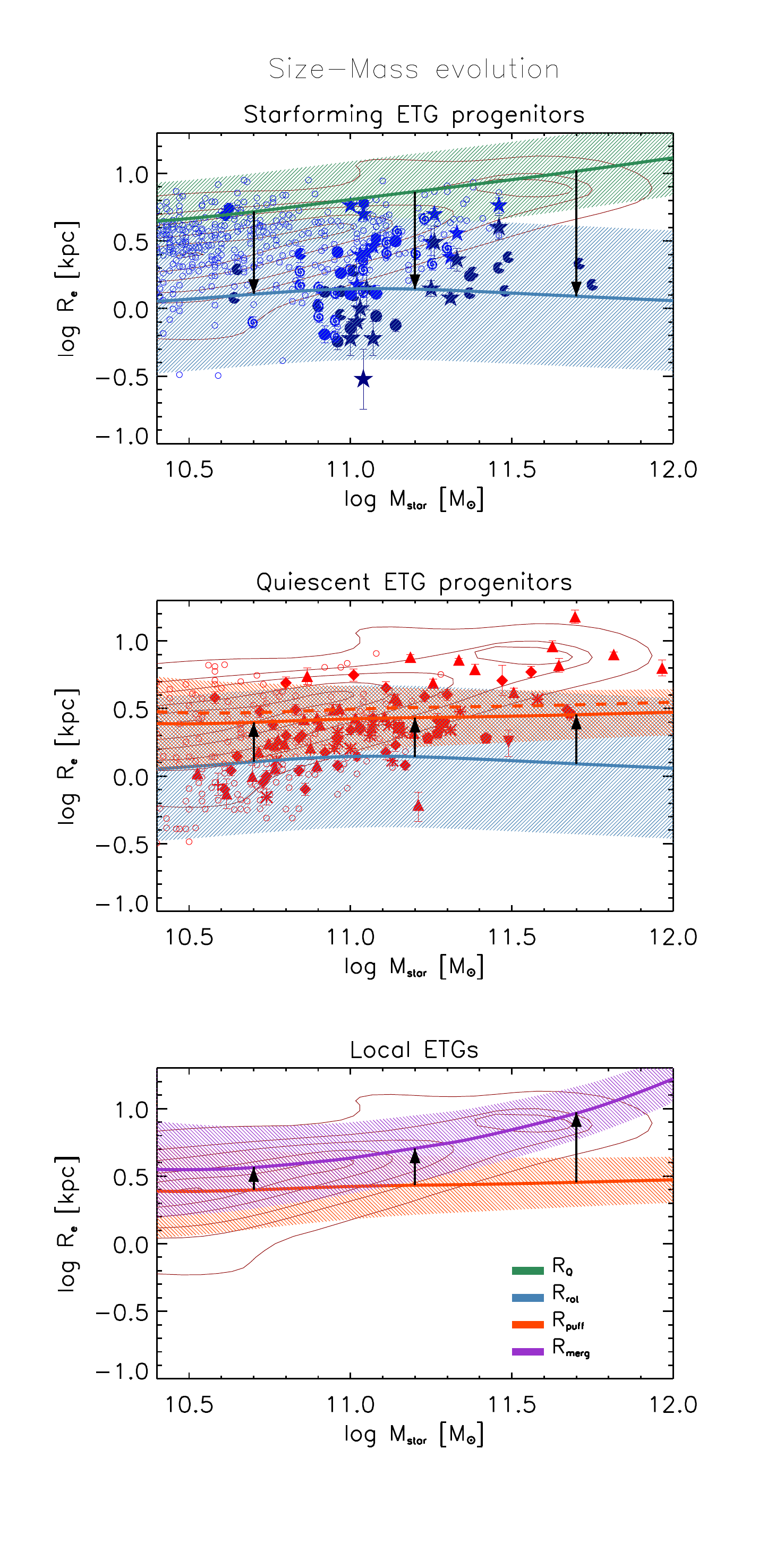}\caption{Dissected evolution of the size $R_e$ vs. stellar mass $M_\star$ relationship for massive ETGs: top panel refers to star-forming ETG progenitors, middle panel to quiescent ETG progenitors, and bottom panel to late-time evolution toward local ETGs. As in previous figure, linestyles and shaded areas refer to relevant sizes (and their associated dispersions), with the black arrows illustrating the evolution from one to another. Colored symbols show various datasets at high-redshift as in previous Figure; contours refer to the local relationship, plotted for reference in each panel.}\label{fig|sizevo}
\end{figure*}

\begin{figure*}
\epsscale{1}\plotone{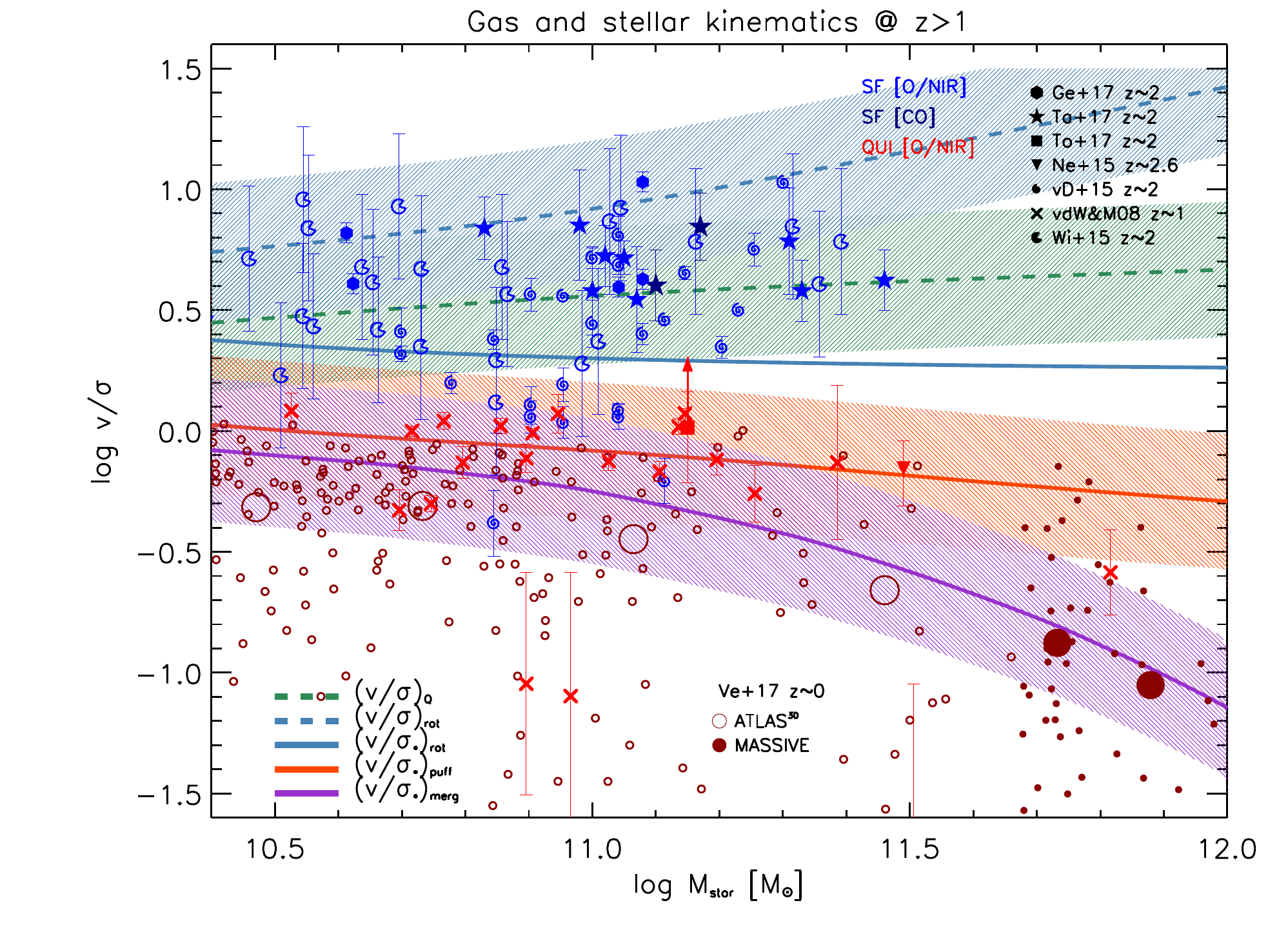}\caption{Ratio $v/\sigma$ between rotation to dispersion velocity vs. stellar mass $M_\star$. The colored lines with shaded areas illustrate the average relationship and scatter expected along the evolution of ETG progenitors: green dashed line refers to the gas velocity ratio $(v/\sigma)_{\rm Q}$ at the fragmentation radius $R_{\rm Q}$, blue dashed line to the gas velocity ratio $(v/\sigma)_{\rm rot}$ at the centrifugal size $R_{\rm rot}$ and blue solid line is the ratio $(v/\sigma)_{\star, \rm rot}$ at the same radius for the stellar component (corresponding shaded area not plotted for clarity), orange solid line to the stellar velocity ratio $(v/\sigma)_{\star, \rm puff}$ at the radius $R_{\rm puff}$ after puffing up, and magenta solid line to the stellar velocity ratio  $(v/\sigma)_{\star, \rm merg}$ at the final radius $R_{\rm merg}$ after dry merging. Data are from Genzel et al. (2017; hexagons) at $z\sim 2$, Tadaki et al. (2017a,b; stars) at $z\sim 2$, Toft et al. (2017; squares) at $z\sim 2$, Newman et al. (2015; reverse triangles) at $z\sim 2.6$, van Dokkum et al. (2015; spirals) at $z\sim 2$, van der Wel \& van der Marel (2008; crosses) at $z\sim 1$, Wisnioski et al. (2015; pacmans) at $z\sim 2$. Blue points refer to star-forming galaxies (light blue for kinematics inferred from optical/near-IR observations, deep blue for kinematics inferred from CO observations), and red points refer to quiescent galaxies. The dark red circles (small circles for individual objects and large circles for the average) illustrate the $(v/\sigma)_\star$ distributions of local ETGs from the ATLAS$^{3\,\rm D}$ (open circles) and MASSIVE (filled circles) surveys as reported by Veale et al. (2017).}\label{fig|vsigma}
\end{figure*}

\begin{figure*}
\epsscale{1}\plotone{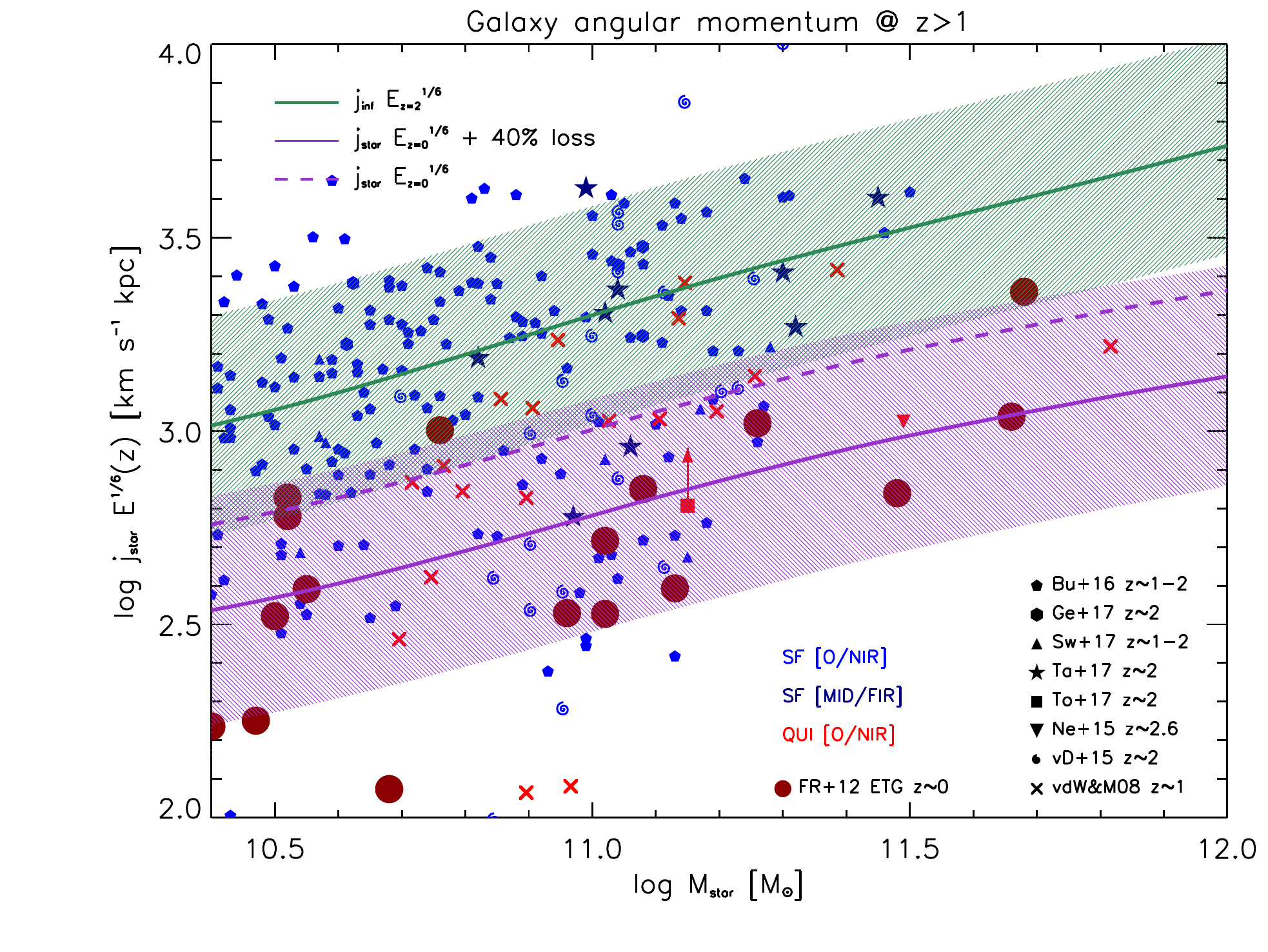}\caption{Specific angular momentum vs. stellar mass. The colored lines with shaded areas illustrate the average relationship and scatter expected along the evolution of ETG progenitors: green line illustrates the specific angular momentum $j_{\rm inf}\, E_{z=2}^{1/6}$ of the infalling gas at $z\sim 2$; magenta lines show the stellar angular momentum $j_\star\,E_{z=0}^{1/6}$ at $z\sim 0$, with the solid line including an angular momentum loss of $40\%$ due to late-time dry mergers and the dashed line referring to no angular momentum loss (only mass addition and redshift scaling included). Data are from Burkert et al. (2016; pentagons) at $z\sim 1-2$, Genzel et al. (2017; hexagons) at $z\sim 2$, Swinbank et al. (2017; triangles) at $z\sim 1-2$, Tadaki et al. (2017a,b; stars) at $z\sim 2$, Toft et al. (2017; squares) at $z\sim 2$, Newman et al. (2015; reverse triangles) at $z\sim 2.6$, van Dokkum et al. (2015; spirals) at $z\sim 2$, van der Wel \& van der Marel (2008; crosses) at $z\sim 1$. Blue points refer to star-forming galaxies (light blue for optical/near-IR observations, deep blue for mid/far-IR observations), and red points refer to quiescent galaxies. The big dark red circles report the angular momentum distribution of local ETGs from Romanowsky \& Fall (2012).}\label{fig|jstar}
\end{figure*}

\begin{itemize}

\item In high-$z$ progenitors of ETGs, the biased collapse of a fraction $f_{\rm inf}\approx 0.4-0.6$ of the baryons initially present in the halo and enclosed within the size $R_{\rm inf}\la 10^2$ kpc sets the timescale $t_{\rm dyn}(R_{\rm inf})\approx$ some $10^8$ yr driving the subsequent evolution. Cooling and fragmentation of the infalling gas occurs on a size $R_{\rm Q}\la 10$ kpc where gaseous clumpy, unstable disk with appreciable rotational motions $(v/\sigma)_{\rm Q}\ga 3$ is formed. Dynamical friction, gravitational torquing and viscosity would imply migration timescales of a few $10^8$ yr for the gas and the clumps; actually in the biased collapse scenario this process in not crucial around $R_{\rm Q}$, since the low specific angular momentum of the gas is not sufficient to sustain the gravitational pull. As a consequence, gas and clumps infall over a dynamical timescale of few tens Myr, approximately maintaining their initial specific angular momentum. During the infall, gas and clumps are expected to feature moderate SFR $\la 50-200\, M_\odot$ yr$^{-1}$, resulting in mild metal enrichment and dust obscuration. As a matter of fact, near-IR/optical observations in $z\sim 1-2$ star-forming galaxies measure size and kinematic properties consistent with our expectations. See Sects.~\ref{sec|sizeform}, \ref{sec|results} and Figs.~\ref{fig|Re_Mstar_ini}, \ref{fig|size}, \ref{fig|sizevo}, \ref{fig|vsigma} for more details.

\item The infall of gas and clumps toward the inner regions halts at around the radius $R_{\rm rot}\la$ 1 kpc where gravity and centrifugal support balance; there the gas kinematics is largely dominated by rotational velocities of several hundreds km s$^{-1}$ corresponding to $v/\sigma\la 10$, and the disk fragmentation is enhanced by the further decrease of the Toomre parameter $Q\la 0.3$. The size $R_{\rm rot}\propto \lambda^2$ depends strongly on the spin parameter $\lambda$ of the host DM halo, and as such features a large dispersion around $0.5$ dex. Further collapse of the gas is made possible by transfer of angular momentum toward the outer regions via dynamical friction over a migration timescale $\la 10^6$ yr. Meanwhile, strong SFRs $\la 500-2000\, M_\odot$ yr$^{-1}$, substantial metal enrichment and dust obscuration are expected to take place within $R_{\rm rot}$. Violent relaxation processes drive the system toward a bulge-like configuration in virial equilibrium; in the end, large stellar masses $M_{\star}\ga$ from several to many $10^{10}\, M_\odot$ are accumulated within $\la 1$ kpc, with appreciable residual rotational support $(v/\sigma)_{\star, \rm rot}\sim 1-2$ in the stellar component. In fact, size and kinematic data from far-IR/sub-mm observations are consistent with our predictions. See Sects.~\ref{sec|sizecomp}, \ref{sec|results} and Figs.~\ref{fig|Re_Mstar_ini}, \ref{fig|size}, \ref{fig|sizevo}, \ref{fig|vsigma} for more details.

\item After several $10^8$ yr, the star-formation process is expected to be quenched by energy feedback from stellar winds/supernovae or, most likely, from the central supermassive BH during its powerful quasar phase: a high-$z$, massive quiescent galaxy is originated. During the quenching, a substantial fraction $f_{\rm out}\approx 0.5-0.7$ of the infalling gas is ejected from the inner regions; this enforces a puffing up of the stellar distribution to a size $R_{\rm puff}\sim 3-5$ kpc, a factor a few to several larger than $R_{\rm rot}$. Even more relevantly, due to the presence of DM, the puffing up process is more pronounced in galaxies that initially were more compact; as a consequence, the dispersion in the size $R_{\rm puff}$ is considerably smaller than that in the initial size $R_{\rm rot}$. Note that as a consequence of the size expansion, the mass concentration in the central region $\la 1$ kpc of quiescent galaxies is somewhat decreased with respect to that of their star-forming progenitors. The puffing up process affects mildly the stellar velocity dispersion, while the rotational velocity is appreciably decreased because of the expansion in size, to yield a velocity ratio $(v/\sigma)_{\star, \rm puff}\ga 0.5$. The current near-IR/optical data on sizes and kinematics of $z\sim 1-2$ massive quiescent galaxies are consistent with such findings. See Sects.~\ref{sec|puffing}, \ref{sec|results} and Figs.~\ref{fig|puffing}, \ref{fig|size}, \ref{fig|sizevo}, \ref{fig|vsigma} for more details.

\item In the subsequent passive evolution toward the present time, mass additions by dry merging events can alter the size and kinematics of quiescent galaxies, and especially so for massive systems. The final size $R_{\rm merg}$ for the most massive galaxies is increased by factor around $2-3$ from $z\sim 2$ to $0$, while the dispersion is somewhat enhanced because of the variance in merging histories.  Meanwhile, the stellar velocity ratio $(v/\sigma)_{\star, \rm merg}$ is decreased somewhat, mainly because the rotation velocity is lowered by the increase in size and by a $40\%$ angular momentum loss via partial spin cancellation during encounters. The outcomes of our analysis are consistent with the observed size $R_e$ and $(v/\sigma)_\star$ distribution of local ETGs, both in average values and dispersion. See Sects.~\ref{sec|merging}, \ref{sec|results} and Figs.~\ref{fig|merging}, \ref{fig|size}, \ref{fig|sizevo}, \ref{fig|vsigma} for more details.

\item We predict, and found agreement in the available data, that the specific angular momentum of ETG progenitors is close to the value dictated by the biased collapse of a fraction $f_{\rm inf}\approx 0.4-0.6$ of the initial baryons. Local ETGs reflects the same momentum but for minor losses due to late-time dry mergers. See Sect.~\ref{sec|size}, \ref{sec|results} and Fig.~\ref{fig|jstar} for more details.

\end{itemize}

It is interesting to compare our findings to the outcomes of recent hydro-cosmological simulations. As to size and kinematics, Zolotov et al. (2015) find that star-forming progenitors of galaxies with final mass $M_\star\sim 10^{11}\, M_{\sun}$ featured small sizes $R_e\la$ kpc and rotational to random velocity ratios $v/\sigma\ga 2$ in the gaseous and $(v/\sigma)_\star\sim 1$ in the stellar component. More recently, both Genel et al. (2018) analysing the \textsl{IllustrisTNG} simulation and Furlong et al. (2017) analysing the \textsl{EAGLE} simulation confirm that the progenitors of local massive quiescent galaxies are characterized by small sizes $\la$ a few kpc during their star-formation phase, while after quenching they experience a substantial size growth due to outward stellar migration, renewed star formation, and mass addition from dry mergers.

A clear prediction of recent simulations is that on the average ETG progenitors evolve toward a low angular momentum state by various processes, such as angular momentum redistribution during compaction, early star-formation quenching by feedbacks, and dry mergers. Zavala et al. (2016) find, via the \textsl{EAGLE} simulations, a strong relation between the specific angular momentum of the stars and that of the host DM halo in the inner star-forming region. Lagos et al. (2017) using the same simulation suite envisage that an early star-formation quenching plus dry mergers can be rather effective in producing galaxies with low specific angular momentum. Both these studies indicate that massive halos with a turnaround epoch $z\ga 2$ typically host central galaxies featuring old stellar populations with low specific angular momentum. On the other hand, the same simulations suggest that halos with a late formation/turnaround epoch $z\la 1$ tend to host disk-dominated galaxies, featuring specific angular momentum in the stars close to that in the overall halo (see Romanowsky \& Fall 2011; Shi et al. 2017; Lapi et al. 2018); the resulting long gas infall timescales and quiet star formation histories yield for disk-dominated galaxies a much less dramatic size and kinematic evolution. Note that in some instances late-time gas recollapse/regrowth around preformed bulges can also contribute to build up disks and to explain the diversity in the bulge-to-disk ratios observed locally (e.g., Bernardi et al. 2014; Moffett et al. 2016).

In a future perspective, the outcomes of our study can provide inspiration toward solving the following hot issues.

\begin{itemize}

\item Improving the (sub-grid) physical recipes implemented in theoretical models and simulations. In particular, the puffing up that follow a substantial gas removal from the inner regions by stellar and/or BH feedback has not yet been included or at least not properly treated in cosmological simulations. Note that in the past years this process has been overlooked by the community, since the short timescale for equilibrium recovery after the puffing vs. the relatively young ages of $z\sim 2$ quiescent galaxies was an argument made against it. The most recent data indicate that, though compact with respect to local ETGs, $z\sim 2$ quiescent galaxies feature sizes significantly larger than compact star-forming objects at similar redshift, so indicating that puffing up has already affected them. Moreover, we have stressed the essential role of the puffing up process in reducing the large dispersion expected and observed in the size of compact star-forming systems. To include the puffing up process in numerical simulations could strongly alleviate problems in reproducing the size distributions of local ETGs and of their progenitors, without adopting extreme feedback or dry mergers prescriptions.

\item Tuning aimed numerical experiments focused on specific processes. Specifically, it would be extremely interesting to elucidate in detail: (i) the role of the biased collapse in setting a low initial specific angular momentum for the gas located in the halo inner regions; (ii) the effectiveness of the compaction via clump/gas migration and outward redistribution of angular momentum vs. feedback processes; (iii) the development of violent relaxation in the inner region to cause a transition from a fully rotation-dominated to a significantly dispersion-endowed configuration; (iv) the possibility that the latter process could also trigger accretion onto the central supermassive BH. Note that current numerical simulations aimed at investigating clump and gas migration often set as initial conditions a rotationally-supported disk with typical size of several kpcs; this means that the initial angular momentum of gas and clumps is quite high, implying migration timescales around a few $10^8$ yr. However, our analysis based on the biased collapse scenario envisages that most of the final stellar component is formed from gas initially characterized by a rather low specific angular momentum; such gas is not rotationally supported  on the fragmentation scale around several kpc, and can infall down to appreciably smaller radii of order kpc maintaining its original angular momentum. The infall halts at the centrifugal barrier, where migration in the innermost regions can occur via dynamical friction over much shorter migration timescales around $10^6$ yr. Numerical test of this scenario requires honed simulations with apt initial conditions, and high space/time resolution.

\item Planning future multi-band, high-resolution observations on high-redshift star-forming/ quiescent galaxies and quasars. For example, it would be very interesting to compare the sizes observed in the far-IR/sub-mm band with that inferred from radio data, to shed light on the spatial scales where radio emission originates in dusty star-forming galaxies; this will be likely become achievable with SKA and its precursors. Other important observations concern kinematic measurements in ETG progenitors, that would constitute crucial test of our scenario. In high-$z$ star-forming galaxies we predict a high $v/\sigma\la 10$ ratio of the gas component in the inner dust-obscured, star-forming regions, in rapid transition toward lower values $\la 2$ via violent relaxation processes; the task is challenging but may be feasible with high-resolution measurements of CO (or other) line profiles by ALMA, especially on gravitationally-lensed objects. On the other hand, in high-$z$ quiescent galaxies we expect a stellar velocity ratio $(v/\sigma)_\star\ga 0.5$; there the measurements at $z\ga 1$ are currently scarce, but their number is expected to increase appreciably in the era of the JWST. Finally, other interesting observations concern extremely high-redshift quasars; a relevant example is the object J1342+0928 at $z\sim 7.5$ studied with JVLA (see Venemans et al. 2017a; Bagnados et al. 2018). Following Eq.~(\ref{eq|tdyn_Rinf}) we expect the infall of about $10^{11}\, M_\odot$ in gas mass (corresponding to the inferred stellar mass of the host, see Venemans et al. 2017a) to occur over a timescale around $10^{8}$ yr, significantly shorter than the Hubble time $\la 700$ Myr at the observed redshift. In addition, based on Eq.~(\ref{eq|Rrot}) we expect a quite small size $\la 0.5$ kpc for the stellar and dust distributions. These specific predictions can be eventually tested by ALMA observations at a resolutions of $\sim 0.1$ arcsec.

\end{itemize}

In summary, we have highlighted the physical mechanisms that in ETG progenitors are responsible for the moderate SFRs $\la 50-200\, M_\odot$ yr$^{-1}$ probed by UV data on scales of several kpcs, for the much higher SFRs $\la 500-2000\, M_\odot$ yr$^{-1}$ probed by far-IR data on (sub-)kpc size, and for the resulting stellar mass growth $M_\star\sim 10^{11}\, M_\odot$ over timescales of some $10^8$ yr on spatial scales of a few to several kpcs, as probed by near-IR observations of quiescent galaxies. The corresponding number densities of UV-selected vs. far-IR selected star-forming galaxies vs. quiescent galaxies have been quantitatively computed via the continuity equation and positively compared with the observed statistics by Lapi et al. (2017b). The dramatic size increase from compact star-forming toward quiescent galaxies at similar redshift is an additional manifestation of the BH-galaxy coevolution, adding to the $\alpha$-enhancement and to the massive outflows detected in quasar hosts at high redshift. To follow the driving processes via numerical simulations, time and spatial resolution should be privileged over large volumes, because of the quite short time- and small length-scales involved. In fact, many observational and theoretical aspects of our analysis suggest a biased collapse scenario for ETG formation, envisaging that the low specific angular momentum of local massive ellipticals has been essentially imprinted since the very beginning, and that in situ processes are more relevant than mergers in driving most of their stellar and BH mass growth.

\begin{acknowledgements}
We thank the referee for stimulating and constructive comments. We are grateful to F. Buitrago, F. Fraternali, M. Negrello, and P. Salucci for helpful discussions, and to J. Miller for critical reading. Work partially supported by PRIN MIUR 2015 `Cosmology and Fundamental Physics: illuminating the Dark Universe with Euclid' and PRIN INAF 2014 `Probing the AGN/galaxy co-evolution through ultra-deep and ultra-high-resolution radio surveys'. AL acknowledges the RADIOFOREGROUNDS grant (COMPET-05-2015, agreement number 687312) of the European Union Horizon 2020 research and innovation programme, and the MIUR grant `Finanziamento annuale individuale attivit\'a base di ricerca'.
\end{acknowledgements}

\begin{appendix}

\section{Mass contrasts}

In this Appendix we provide explicit computation of the mass contrasts at the relevant size scales used in the main text. The critical radius $R_{\rm Q}$ for clump fragmentation is given by the equation
\begin{equation}
R_{\rm  Q} = {j_{\rm inf}\, Q\over \sqrt{2}\,\sigma}\, \delta_{\rm gas}(R_{\rm Q})
\end{equation}
where the gas mass contrast is defined as $\delta_{\rm gas}(R_{\rm Q})\equiv M_{\rm gas}(<R_{\rm Q})/M_{\rm tot}(<R_{\rm Q})$.

We assume that the baryonic mass within $R_{\rm Q}$ is constituted by all the infalling mass $M_{\rm inf}=f_{\rm inf}\, f_b\,M_{\rm H}$, and that the gas mass can be estimated as $M_{\rm gas}\simeq M_{\rm inf}-M_\star$; this implies that the gas fraction reads $f_{\rm gas}\equiv M_{\rm gas}/M_{\rm inf}\approx 1-f_\star/f_{\rm inf}$, with typical values around $\ga 0.5$ as observed in high-redshift disks (see Tacconi et al. 2013, 2018; Saintonge et al. 2013; Genzel et al. 2015; Barro et al. 2017). Note that actually most of this gas mass will be then evacuated from the galaxy after a Gyr, by feedback from the central supermassive BH/quasar (see Sect.~\ref{sec|puffing}). As to the DM mass, it can be written as $M_{H}(<R_{\rm Q})= 0.1\, M_{\rm H}\, (R_{\rm Q}/0.1\, R_H)^2$, since $M_{H}(<R)\propto R$ for $R\ga 0.1\, R_{\rm H}$ and $M_{H}(<R)\propto R^2$ for $R\la 0.1\, R_{\rm H}$ approximately hold; a posteriori one can check that $R_{\rm Q}$ falls in the latter radial range. All in all, the gas mass contrast is given by
\begin{equation}
\delta_{\rm gas}(R_{\rm Q}) \simeq {f_{\rm gas}\,f_{\rm inf}\, f_b\,M_{\rm H}\over f_{\rm inf}\, f_b\, M_{\rm H}+0.1\, M_{\rm H}\, (R_{\rm Q}/0.1\,R_{\rm H})^2} = {f_{\rm gas}\over 1+x_{\rm Q}^2/0.1\, f_{\rm inf}\,f_b}~,
\end{equation}
where $x_{\rm Q}\equiv R_{\rm Q}/R_{\rm H}$. Thus using Eq.~(A1) yields the implicit equation
\begin{equation}
x_{\rm Q} \approx {j_{\rm inf}\, Q\over \sqrt{2}\,\sigma\, R_{\rm H}}\, {f_{\rm gas}\over 1+x_{\rm Q}^2/0.1\, f_{\rm inf}\, f_b}~.
\end{equation}
Using the reference values $j_{\rm inf}\approx 1.4\times 10^3$ km s$^{-1}$ kpc, $f_{\rm inf}\approx 0.6$, $f_{\star}\approx 0.2$, $Q\approx 1$ and $\sigma\approx 60$ km s$^{-1}$ applying for $M_\star\approx 10^{11}\, M_\odot$, the numerical solution yields $x_{\rm Q}\approx 0.05$ corresponding to $\delta_{\rm gas}(R_{\rm Q})\approx 0.5$. We stress that to obtain the quantitative results on $R_{\rm Q}$ presented in the figures of the main text, the above computation of $\delta_{\rm gas}(R_{\rm Q})$ has been performed with the detailed dependence of the parameters on the stellar mass.

So far we have neglected the adiabatic contraction of the DM component; now we will take it into account with an iterative scheme. The classic equation to describe the process is
\begin{equation}
r_f\, [M_D(<r_f) + M_i(<r_i)\, (1-m_D)] = r_i\, M_i(<r_i)
\end{equation}
where $r_i$ and $r_f$ are the radii before and after contraction, $M_D(<r_i)$ is the mass in the disk within $r_i$, $m_D$ is the fraction of mass in the disk, $M_i(<r_i)$ is the DM mass within $r_i$. In the present context we can identify $r_f\rightarrow R_{\rm Q}$ where $R_{\rm Q}$ is the known solution without adiabatic contraction, $r_i\rightarrow \tilde R_{\rm Q}$ where $\tilde R_{\rm Q}$ is the unknown starting radius with adiabatic contraction, $M_D(<r_f)\equiv M_\star=f_\star\, f_b\,M_{\rm H}$, $m_D\equiv f_\star\,f_b$, and $M_i(<r_i)\equiv 0.1\, M_{\rm H}\, (\tilde R_{\rm Q}/0.1\, R_{\rm H})^2$. Then Eq.~(A4) becomes
\begin{equation}
R_{\rm Q}\, \left[M_\star+(1-f_\star\, f_b)\, 0.1\, M_{\rm H}\, \left({\tilde R_{\rm Q}\over 0.1\, R_{\rm H}}\right)^2\right] = \tilde R_{\rm Q}\, 0.1\, M_{\rm H}\, \left({\tilde R_{\rm Q}\over 0.1\, R_{\rm H}}\right)^2~,
\end{equation}
and defining $\tilde x_{\rm Q}\equiv \tilde R_{\rm Q}/R_{\rm H}$ leads to the algebraic equation
\begin{equation}
x_{\rm Q}\, [0.1\,f_\star\, f_b+(1-f_\star\, f_b)\, \tilde x_{\rm Q}^2] = \tilde x_{\rm Q}^3~.
\end{equation}
Using the solution $x_{\rm Q}\approx 0.05$ from Eq.~(A3) yields $\tilde x_{\rm Q}\approx 0.08$ corresponding to $\delta_{\rm gas}(R_{\rm Q})\approx 0.38$. In addition, the baryonic contrast $\delta(R_{\rm Q})\equiv M_{\rm inf}/M_{\rm tot}(<R_{\rm Q})$ involving all the infalling mass is given by
\begin{equation}
\delta(R_{\rm Q})\simeq {f_{\rm inf}\, f_b\,M_{\rm H}\over f_{\rm inf}\, f_b\, M_{\rm H}+0.1\, M_{\rm H}\, (R_{\rm Q}/0.1\,R_{\rm H})^2} = {1\over 1+x_{\rm Q}^2/0.1\, f_{\rm inf}\,f_b}~,
\end{equation}
and takes on values around $\delta(R_{\rm Q})\approx 0.6$.

The same line of reasoning can be applied to the computation of the radius
$R_{\rm rot}$ of the centrifugal barrier, given by
\begin{equation}
R_{\rm rot} = {j_{\rm inf}^2\over G\,M_{\rm inf}}\, \delta(R_{\rm rot})
\end{equation}
where the baryonic mass contrast is now defined as $\delta(R_{\rm rot})\equiv M_{\rm inf}/M_{\rm tot}(<R_{\rm rot})$. We derive
\begin{equation}
\delta(R_{\rm rot}) \approx {1\over 1+x_{\rm rot}^2/0.1\, f_{\rm inf}\,f_b}~,
\end{equation}
where $x_{\rm rot}\equiv R_{\rm rot}/R_{\rm H}$ is given by the implicit equation
\begin{equation}
x_{\rm rot} \approx {j_{\rm inf}^2\over G\, M_{\rm inf}\, R_{\rm H}}\, {1\over 1+x_{\rm rot}^2/0.1\, f_{\rm inf}\, f_b}~.
\end{equation}
The numerical solution yields $x_{\rm rot}\approx 0.0099$ corresponding to $\delta(R_{\rm rot})\approx 0.99$.

Introducing now the effects of adiabatic contraction, the corrected radius $\tilde R_{\rm rot}\equiv \tilde x_{\rm rot}\, R_{\rm H}$ is determined by
\begin{equation}
x_{\rm rot}\, [0.1\,f_\star\, f_b+(1-f_\star\, f_b)\, \tilde x_{\rm rot}^2] = \tilde x_{\rm rot}^3~,
\end{equation}
which yields $\tilde x_{\rm rot}\approx 0.036$ corresponding to $\delta(R_{\rm rot})\approx 0.88$. In addition, the gas mass contrast $\delta_{\rm gas}(R_{\rm rot})\equiv M_{\rm gas}/M_{\rm tot}(<R_{\rm rot})$ is given by
\begin{equation}
\delta_{\rm gas}(R_{\rm rot})\simeq {f_{\rm gas}\, f_{\rm inf}\, f_b\,M_{\rm H}\over f_{\rm inf}\, f_b\, M_{\rm H}+0.1\, M_{\rm H}\, (R_{\rm rot}/0.1\,R_{\rm H})^2} = {f_{\rm gas}\over 1+x_{\rm rot}^2/0.1\, f_{\rm inf}\,f_b}~,
\end{equation}
and takes on values around $\delta_{\rm gas}(R_{\rm rot})\approx 0.57$.

Finally, we aim at computing the radius $R_b$ where the baryonic mass dominate the gravitational potential over the DM. To first approximation this is given by
\begin{equation}
{G\, M_{\rm inf}\over R_{\rm b}} \approx {G\, M_{\rm H}(<R_{\rm b})\over R_{\rm b}}~,
\end{equation}
corresponding to a baryonic mass contrast $\delta(R_{\rm b})\equiv M_{\rm inf}/M_{\rm tot}(<R_{\rm b})\approx 0.5$. Neglecting adiabatic contraction and assuming the scaling $M_{\rm H}(<R_{\rm b})\simeq M_{\rm H}\, x_{\rm b}^2/0.1$ with $x_{\rm b}\equiv R_{\rm b}/R_{\rm H}$ yields $x_{\rm b}\simeq \sqrt{0.1\,f_{\rm inf}\, f_b}\approx 0.098$.

Introducing now the effects of adiabatic contraction, the corrected radius $\tilde R_{\rm b}\equiv \tilde x_{\rm b}\, R_{\rm H}$ is determined by
\begin{equation}
x_{\rm b}\, [0.1\,f_\star\, f_b+(1-f_\star\, f_b)\, \tilde x_{\rm b}^2] = \tilde x_{\rm b}^3~,
\end{equation}
which yields $\tilde x_b\approx 0.12$, corresponding to a mass contrast $\delta(R_{\rm b})\approx 0.41$. All in all, the radius $R_{\rm b}$ writes
\begin{equation}
R_{\rm b} \approx \sqrt{[\delta(R_{\rm b})^{-1}-1]\, 0.1\, f_{\rm inf}\, f_b}\, R_{\rm H}~.
\end{equation}

\end{appendix}

\clearpage

\clearpage

\clearpage

\clearpage

\clearpage

\clearpage

\clearpage

\clearpage

\clearpage

\begin{references}

\reference{}Alexander, D. M., \& Hickox, R. C. 2012, NewAR, 56, 93

\reference{}Andrews, B. H., \& Thompson, T. A. 2011, ApJ, 727, 97

\reference{}Annibali, F., Bressan, A., Rampazzo, R., Zeilinger, W. W., \& Danese, L. 2007, A\&A, 463, 455

\reference{}Aravena, M., Spilker, J. S., Bethermin, M., et al. 2016, MNRAS, 457, 4406

\reference{}Aversa, R., Lapi, A., De Zotti, G., Shankar, F., \& Danese, L. 2015, ApJ, 810, 74

\reference{}Ba\~{n}ados, E., Venemans, B.P., Mazzucchelli, C., et al. 2018, Natur, 553, 473

\reference{}Barnes, J., \& Efstathiou, G. 1987, ApJ, 319, 575

\reference{}Barro, G., Kriek, M., Perez-Gonzalez, P. G., et al. 2017, ApJ, 851, L40

\reference{}Barro, G., Kriek, M., P\'erez-Gonz\'alez, P. G., et al. 2016a, ApJ, 827, L32

\reference{}Barro, G., Faber, S.M., Dekel, A., et al. 2016b, ApJ, 820, 120

\reference{}Barro, G., Trump, J.R., Koo, D.C., et al. 2014, ApJ, 795, 145

\reference{}Baumgardt, H., \& Kroupa, P. 2007, MNRAS, 380, 1589

\reference{}Beckmann, R. S., Devriendt, J., Slyz, A., et al. 2017, MNRAS, 472, 949

\reference{}Behroozi, P. S., Wechsler, R. H., \& Conroy, C. 2013, ApJ, 770, 57

\reference{}Belli, S., Newman, A.B., \& Ellis, R.S. 2017, ApJ, 834, 18

\reference{}Belli, S., Newman, A.B., Ellis, R.S., \& Konidaris, N.P. 2014, ApJ, 788, L29

\reference{}Bernardi, M., Meert, A., Vikram, V. et al. 2014, MNRAS, 443, 874

\reference{}B\'ethermin, M., Daddi, E., Magdis, G., et al. 2015, A\&A, 573, A113

\reference{}Biermann, P., \& Shapiro, S. L. 1979, ApJ, 230, L33

\reference{}Binney, J. 2005, MNRAS, 363, 937

\reference{}Birnboim, Y., \& Dekel, A. 2003, MNRAS, 345, 349

\reference{}Boily, C. M., \& Kroupa, P. 2003, MNRAS, 338, 673

\reference{}Bournaud, F. 2016, in Galactic Bulges, Astrophysics and Space Science Library vol. 418 (Switzerland: Springer International Publishing), p. 355

\reference{}Bournaud, F., Perret, V., Renaud, F., et al. 2014, ApJ, 780, 57

\reference{}Bournaud, F., Chapon, D., Teyssier, R., et al. 2011, ApJ, 730, 4

\reference{}Bourne, N., Dunlop, J. S., Merlin, E., et al. 2017, MNRAS, 467, 1360

\reference{}Bouwens, R. J., Oesch, P. A., Illingworth, G. D., Ellis, R. S., \& Stefanon, M. 2017, ApJ, 843, 129

\reference{}Bouwens, R. J., Aravena, M., De Carli, R., et al. 2016, ApJ, 833, 72

\reference{}Boylan-Kolchin, M., Ma, C.-P., Quataert, E. 2008, MNRAS, 383, 93

\reference{}Braun, H., \& Schmidt, W. 2012, MNRAS, 421, 1838

\reference{}Bressan, A., Silva, L., \& Granato, G. L. 2002, A\&A, 392, 377

\reference{}Buitrago, F., Trujillo, I., Curtis-Lake, E., et al. 2017, MNRAS, 466, 4888

\reference{}Bullock, J. S., Dekel, A., Kolatt, T. S., et al. 2001, ApJ, 555, 240

\reference{}Burkert, A., Forster Schreiber, N. M., Genzel, R., et al. 2016, ApJ, 826, 214

\reference{}Cappellari, M. 2016, ARA\&A, 54, 597

\reference{}Cappellari, M., McDermid, R.M., Alatalo, K., et al. 2013, MNRAS, 432, 1862

\reference{}Caputi, K. I., Ilbert, O., Laigle, C., et al. 2015, ApJ, 810, 73

\reference{}Carniani, S., Marconi, A., Maiolino, R., et al. 2017, A\&A, 605, A105

\reference{}Ceverino, D., Dekel, A., \& Bournaud, F. 2010, MNRAS, 404, 2151

\reference{}Chabrier, G. 2003, ApJL, 586, L133

\reference{}Chartas, G., Saez, C., Brandt, W. N., Giustini, M., \& Garmire, G. P. 2009, ApJ, 706, 644

\reference{}Cimatti, A., Cassata, P., Pozzetti, L., et al. 2008, A\&A, 482, 21

\reference{}Ciotti, L., \& Ostriker, J. P. 2007, ApJ, 665, 1038

\reference{}Choi, J., Conroy, C., Moustakas, J., et al. 2014, ApJ, 792, 95

\reference{}Citro, A., Pozzetti, L., Moresco, M., \& Cimatti, A. 2016, A\&A, 592, A19

\reference{}Cole, S., Lacey, C.G., Baugh, C.M., \& Frenk, C.S. 2000, MNRAS, 319, 168

\reference{}Cooray, A., Calanog, J., Wardlow, J. L., et al. 2014, ApJ, 790, 40

\reference{}Damjanov, I., McCarthy, P.J., Abraham, R.G., et al. 2009, ApJ, 695, 101

\reference{}Danovich, M., Dekel, A., Hahn, O., Ceverino, D., \& Primack, J. 2015, MNRAS, 449, 2087

\reference{}Davidzon, I., Ilbert, O., Laigle, C., et al. 2017, A\&A, 605, A70

\reference{}Decarli, R., Walter, F., Venemans, B. P., et al. 2017, Natur, 545, 457

\reference{}Decarli, R., Walter, F., Aravena, M., et al. 2016, ApJ, 833, 70

\reference{}DeGraf, C., Dekel, A., Gabor, J., \& Bournaud, F. 2017, MNRAS, 466, 1462

\reference{}Dekel, A., \& Burkert, A. 2014, MNRAS, 438, 1870

\reference{}Dekel, A. Birnboim, Y., Engel, G., et al., 2009, Natur, 457, 451

\reference{}Dekel, A., \& Birnboim, Y. 2008, MNRAS, 383, 119

\reference{}Dekel, A., \& Krumholz, M.R. 2013, MNRAS, 432, 455

\reference{}Delvecchio, I., Lutz, D., Berta, S., et al. 2015, MNRAS, 449, 373

\reference{}Di Teodoro, E. M., Fraternali, F., \& Miller, S. H. 2016, A\&A, 594, A77

\reference{}D'Onghia, E., \& Burkert, A. 2004, ApJ, 612, L13

\reference{}Duncan, K., Conselice, C. J., Mortlock, A., et al. 2014, MNRAS, 444, 2960

\reference{}Dunlop, J. S., McLure, R. J., Biggs, A. D., et al. 2017, MNRAS, 466, 861

\reference{}Dutton, A.A., Conroy, C., van den Bosch, F.C., et al. 2011, MNRAS, 416, 322

\reference{}Eke, V., Efstathiou, G., \& Wright, L. 2000, MNRAS, 315, L18

\reference{}Elmegreen, D.M., Elmegreen, B. G., Ravindranath, S., \& Coe, D. A. 2007, ApJ, 658, 763

\reference{}Elmegreen, D. M., Elmegreen, B. G., \& Ferguson, T. E. 2005, ApJ, 623, L71

\reference{}Emsellem, E., Cappellari, M., Krajnovic, D., et al. 2007, MNRAS, 379, 401

\reference{}Faber, S. M., \& Jackson, R. E. 1976, ApJ, 204, 668

\reference{}Fall, S. M. 2002, in ASP Conf. Ser. 275, Disks of Galaxies: Kinematics, Dynamics and Peturbations, ed. E. Athanassoula, A. Bosma, \& R. Mujica (San Francisco, CA: ASP), 389

\reference{}Fan, L., Lapi, A., Bressan, A., et al. 2010, ApJ, 718, 1460

\reference{}Fan, L., Lapi, A., De Zotti, G., \& Danese, L. 2008, ApJL, 689, L101

\reference{}Feldmann, R. 2015, MNRAS, 449, 3274

\reference{}Finlator, K., Oh, S. P., Ozel, F., \& Dav\'e, R. 2012, MNRAS, 427, 2464

\reference{}Fiore, F., Feruglio, C., Shankar, F., et al. 2017, A\&A, 601, A143

\reference{}Furlong, M., Bower, R. G., Crain, R. A., et al. 2017, MNRAS, 465, 722

\reference{}Gabor, J. M., \& Bournaud, F. 2013, MNRAS, 434, 606

\reference{}Gallazzi, A., Bell, E. F., Zibetti, S., Brinchmann, J., \& Kelson, D. D. 2014, ApJ, 788, 72

\reference{}Gallazzi, A., Charlot, S., Brinchmann, J., \& White, S. D. M. 2006, MNRAS, 370, 1106

\reference{}Genel, S., Nelson, D., Pillepich, A., et al. 2018, MNRAS, 474, 3976

\reference{}Genzel, R., Forster Schreiber, N. M., Ubler, H., et al. 2017, Natur, 543, 397

\reference{}Genzel, R., Tacconi, L. J., Lutz, D., et al. 2015, ApJ, 800, 20

\reference{}Genzel, R.,  Forster Schreiber, N. M., Lang, P., et al. 2014, ApJ, 785, 75

\reference{}Genzel, R., Newman, S., Jones, T., et al. 2011, ApJ, 733, 101

\reference{}Geyer, M. P., \& Burkert, A. 2001, MNRAS, 323, 988

\reference{}Glazebrook, K., Schreiber, C., Labb\'e, I., et al. 2017, Natur, 544, 71

\reference{}Goldreich, P., \& Tremaine, S. 1980, ApJ, 241, 425

\reference{}Gomez-Guijarro, C., Toft, S., Karim, A., et al. 2018, ApJ, in press [arXiv:1802.07751]

\reference{}Gonzalez, A.H., Sivanandam, S., Zabludoff, A.I., \& Zaritsky, D. 2013, ApJ, 778, 14

\reference{}Goodwin, S. P., \& Bastian, N. 2006, MNRAS, 373, 752

\reference{}Granato, G. L., De Zotti, G., Silva, L., Bressan, A., \& Danese, L. 2004, ApJ, 600, 580

\reference{}Grazian, A., Fontana, A., Santini, P., et al. 2015, A\&A, 575, A96

\reference{}Gruppioni, C., Calura, F., Pozzi, F., et al. 2015, MNRAS, 451, 3419

\reference{}Gruppioni, C., Pozzi, F., Rodighiero, G., et al. 2013, MNRAS, 432, 23

\reference{}Guo, Y., Rafelski, M., Bell, E.F., et al. 2018, ApJ, 853, 108

\reference{}Hill, R., Chapman, S.C., Scott, D., et al. 2017, MNRAS, in press [arXiv:1710.02231]

\reference{}Hills, J. G. 1980, ApJ, 235, 986

\reference{}Hodge, J. A., Swinbank, A. M., Simpson, J. M., et al. 2016, ApJ, 833, 103

\reference{}Hodge, J. A., Karim, A., Smail, I., et al. 2013, ApJ, 768, 91

\reference{}Hopkins, P.F., Quataert, E., \& Murray, N. 2012, MNRAS, 421, 3522

\reference{}Hopkins, P. F., Hernquist, L., Cox, T. J., Robertson, B., \& Springel, V. 2006, ApJS, 163, 50

\reference{}Hudson, M. J., Gillis, B. R., Coupon, J., et al. 2015, MNRAS, 447, 298

\reference{}Huertas-Company, M., Perez-Gonzalez, P. G., Mei, S., et al. 2015, ApJ, 809, 95

\reference{}Huynh, M.T., Emonts, B. H. C., Kimball, A. E., et al. 2017, MNRAS, 467, 1222

\reference{}Hyde, J. B., \& Bernardi, M. 2009, MNRAS, 394, 1978

\reference{}Ikarashi, S., Ivison, R. J., Caputi, K. I., et al. 2017, ApJ, 835, 286

\reference{}Ikarashi, S., Ivison, R. J., Caputi, K.I., et al. 2015, ApJ, 810, 133

\reference{}Ilbert, O., McCracken, H. J., le Fevre, O., et al. 2013, A\&A, 556, A55

\reference{}Iliev, I. T., Mellema, G., Shapiro, P. R., \& Pen, U.-L. 2007, MNRAS, 376, 534

\reference{}Immeli, A., Samland, M., Westera, P., \& Gerhard, O. 2004, ApJ, 611, 20

\reference{}Jimenez-Andrade, E. F., Magnelli, B., Karim, A., et al. 2018, A\&A, in press [arXiv:1710.10181]

\reference{}Johnson, H. L., Harrison, C. M., Swinbank, A. M., et al. 2018, MNRAS, 474, 5076

\reference{}Johnson, J. L., Whalen, D. J., Li, H., \& Holz, D. E. 2013, ApJ, 771, 116

\reference{}Karim, A., Swinbank, A. M., Hodge, J. A., et al. 2013, MNRAS, 432, 2

\reference{}Khochfar, S., \& Ostriker, J. P. 2008, ApJ, 680, 54

\reference{}Khochfar, S., \& Silk, J. 2006, ApJ, 648, L21

\reference{}Koprowski, M., Dunlop, J. S., Michalowski, M. J., et al. 2016, MNRAS, 458, 4321

\reference{}Koprowski, M. P., Dunlop, J. S., Michalowski, M. J., Cirasuolo, M., \& Bowler, R. A. A. 2014, MNRAS, 444, 117

\reference{}Kormendy, J., \& Ho, L. C. 2013, ARA\&A, 51, 511

\reference{}Kravtsov, A., Vikhlinin, A., \& Meshscheryakov, A.
2014 [arXiv:1401.7329]

\reference{}Kriek, M., Conroy, C., van Dokkum, P.G., et al. 2016, Natur, 540, 248

\reference{}Krumholz, M.R., Dekel, A., \& McKee, C.F. 2012, 745, 69

\reference{}Krumholz, M. R., \& Dekel, A. 2012, ApJ, 753, 16

\reference{}Kurczynski, P., Gawiser, E., Acquaviva, V., et al. 2016, ApJL, 820, L1

\reference{}Lagos, C. d. P., Theuns, T., Stevens, A. R. H., et al. 2017, MNRAS, 464, 3850

\reference{}Lang, P., Forster Schreiber, N.M., Genzel, R. 2017, ApJ, 840, 92

\reference{}Lange, R., Driver, S. P., Robotham, A.S.G. 2015, MNRAS, 447, 2603

\reference{}Lapi, A., Salucci, P., \& Danese, L. 2018, ApJ, submitted

\reference{}Lapi, A., Mancuso, C., Celotti, A., \& Danese, L. 2017a, ApJ, 835, 37

\reference{}Lapi, A., Mancuso, C., Bressan, A., \& Danese, L. 2017b, ApJ, 847, 13

\reference{}Lapi, A., Raimundo, S., Aversa, R., et al. 2014, ApJ, 782, 69

\reference{}Lapi, A., Gonzalez-Nuevo, J., Fan, L., et al. 2011, ApJ, 742, 24

\reference{}Law, D. R., Steidel, C.C., Erb, D.K., et al. 2009, ApJ, 697, 2057

\reference{}Lilly, S. J., Carollo, C. M., Pipino, A., et al. 2013, ApJ, 772, 119

\reference{}Lonoce, I., Longhetti, M., Maraston, C., et al. 2015, MNRAS, 454, 3912

\reference{}Macci\'o, A.V., Dutton, A.A., van den Bosch, F.C., Moore, B., Potter, D., \& Stadel, J. 2007, MNRAS, 378, 55

\reference{}Maller, A.H., Dekel, A., \& Somerville, R. 2002, MNRAS, 329, 423

\reference{}Man, A. W. S., Greve, T.R., Toft, S., et al. 2016, ApJ, 820, 11

\reference{}Mancuso, C., Lapi, A., Shi, J., et al. 2016a, ApJ, 823, 128

\reference{}Mancuso, C., Lapi, A., Shi, J., et al. 2016b, ApJ, 833, 152

\reference{}Mandelbaum, R., Wang, W., Zu, Y., et al. 2016, MNRAS, 457, 3200

\reference{}Mandelker, N., Dekel, A., Ceverino, D., DeGraf, C., Guo, Y., \& Primack, J. 2017, MNRAS, 464, 635

\reference{}Mandelker, N., Dekel, A., Ceverino, D., Tweed, D.,
Moody, C. E., \& Primack, J. 2014, MNRAS, 443, 3675

\reference{}Martin-Navarro, I., Vazdekis, A., Falcon-Barroso, J., La Barbera, F., Yildirim, A., \& van de Ven, G. 2018, MNRAS, 475, 3700

\reference{}Massardi, M., Enia, A.F.M., Negrello, M., et al. 2018, A\&A, 610, A53

\reference{}Mihos, J. C., \& Hernquist, L. 1996, ApJ, 464, 641

\reference{}Mo, H. J., Mao, S., \& White, S. D. M. 1998, MNRAS, 295, 319

\reference{}Mo, H., van den Bosch, F. C., \& White, S. 2010, Galaxy Formation and Evolution (Cambridge: Cambridge Univ. Press)

\reference{}Moffett, A.J., Lange, R., Driver, S.P., et al. 2016, MNRAS, 462 , 4336

\reference{}More, S., van den Bosch, F. C., Cacciato, M., et al. 2011, MNRAS, 410, 210

\reference{}Moster, B.P., Naab, T., \& White, S. D. M. 2017, MNRAS, submitted [arXiv:1705.05373]

\reference{}Moster, B. P., Naab, T., \& White, S. D. M. 2013, MNRAS, 428, 3121

\reference{}Mullaney, J. R., Daddi, E., Bethermin, M., et al. 2012, ApJL, 753, L30

\reference{}Murray, N., Quataert, E., \& Thompson, T. A. 2010, ApJ, 709, 191

\reference{}Murray, N., Quataert, E., \& Thompson, T. A. 2005, ApJ, 618, 569

\reference{}Naab, T., Johansson, P.H., \& Ostriker, J.P. 2009, ApJ, 699, L178

\reference{}Narayanan, D., Turk, M., Feldmann, R., et al. 2015, Natur, 525, 496

\reference{}Nayyeri, H., Keele, M., Cooray, A., et al. 2016, ApJ, 823, 17

\reference{}Navarro, J. F., Frenk, C. S., \& White, S. D. M. 1997, ApJ, 490, 493

\reference{}Negrello, M., Amber, S., Amvrosiadis, A., et al. 2017, MNRAS, 465, 3558

\reference{}Negrello, M., Hopwood, R., Dye, S., et al. 2014, MNRAS, 440, 1999

\reference{}Newman, A.B., Belli, S., \& Ellis, R.S. 2015, ApJ, 813, L7

\reference{}Noguchi M. 1999, ApJ, 514, 77

\reference{}Nomoto, K., Kobayashi, C., \& Tominaga, N. 2013, ARA\&A, 51, 457

\reference{}Novak, M., Smolcic, V., Delhaize, J., et al. 2017, A\&A, 602, A5

\reference{}Oklopcic, A., Hopkins, P.F., Feldmann, R., Keres, D., Faucher-Giguere, C.-A., \& Murray, N. 2017, MNRAS, 465, 952

\reference{}Page, M. J., Symeonidis, M., Vieira, J., et al. 2012, Natur, 485, 213

\reference{}Pawlik, A. H., Schaye, J., \& van Scherpenzeel, E. 2009, MNRAS, 394, 1812

\reference{}Planck Collaboration XIII 2016, A\&A, 594, A13

\reference{}Pope, A., Montana, A., Battisti, A., et al. 2017, ApJ, 838, 137

\reference{}Prochaska, J. X., \& Hennawi, J. F. 2009, ApJ, 690, 1558

\reference{}Ragone-Figueroa, C., \& Granato, G.L. 2011, MNRAS, 414, 3690

\reference{}Renzini, A. 2006, ARA\&A, 44, 141


\reference{}Richstone, D. O., \& Potter, M. D. 1982, ApJ, 254, 451

\reference{}Riechers, D. A., Daisy Leung, T. K., Ivison, R., et al. 2017, ApJ, 850, 1

\reference{}Rodighiero, G., Brusa, M, Daddi, E., et al. 2015, ApJ, 800, L10

\reference{}Rodighiero, G., Daddi, E., Baronchelli, I., et al. 2011, ApJ, 739, L40

\reference{}Rodriguez-Gomez, V., Pillepich, A., Sales, L.V., et al. 2016, MNRAS, 458, 2371

\reference{}Rodriguez-Gomez, V., Genel, S., Vogelsberger, M., et al. 2015, MNRAS, 449, 49

\reference{}Rodriguez-Puebla, A., Avila-Reese, V., Yang, X., et al. 2015, ApJ, 799, 130

\reference{}Romano, D., Karakas, A. I., Tosi, M., \& Matteucci, F. 2010, A\&A, 522, A32

\reference{}Romanowsky, A. J., \& Fall, S. M. 2012, ApJS, 203, 17

\reference{}Rowan-Robinson, M., Oliver, S., Wang, L., et al. 2016, MNRAS, 461, 1100

\reference{}Rujopakarn, W., Nyland, K., Rieke, G. H., et al. 2018, ApJ, 854, L4

\reference{}Saintonge, A., Lutz, D., Genzel, R., et al. 2013, ApJ, 778, 2

\reference{}Salmon, B., Papovich, C., Finkelstein, S. L., et al. 2015, ApJ, 799, 183

\reference{}Santini, P., Fontana, A., Castellano, M., et al. 2017, ApJ, 847, 76

\reference{}Schreiber, C., Labb\'e, I., Glazebrook, K., et al. 2017, A\&A, in press [arXiv:1709.03505]

\reference{}Schreiber, C., Pannella, M., Leiton, R., et al. 2017b, A\&A, 599, A134

\reference{}Scoville, N., Lee, N., Vanden Bout, P., et al.  2017, ApJ, 837, 150

\reference{}Scoville, N., Sheth, K., Aussel, H., et al. 2016, ApJ, 820, 83

\reference{}Scoville, N., Aussel, H., Sheth, K., et al. 2014, ApJ, 783, 84

\reference{}Shankar, F., Bernardi, M., Sheth, R.K., et al. 2016, MNRAS, 460, 3119

\reference{}Shankar, F., Mei, S., Huertas-Company, M., et al. 2014, MNRAS, 439, 3189

\reference{}Shankar, F., Marulli, F., Bernardi, M., et al. 2013, MNRAS, 428, 109

\reference{}Shankar, F., Bernardi, M., \& Haiman, Z. 2009, ApJ, 694, 867

\reference{}Shankar, F., Lapi, A., Salucci, P., de Zotti, G., \& Danese, L. 2006, ApJ, 643, 14

\reference{}Shen, S., Mo, H.J., White S.D.M., et al. 2003, MNRAS, 343, 978

\reference{}Shi, J., Lapi, A., Mancuso, C., Wang, H., \& Danese, L. 2017, ApJ, 843, 105

\reference{}Shibuya, T., Ouchi, M., \& Harikane, Y. 2015, ApJS, 219, 15

\reference{}Shlosman, I., \& Noguchi, M. 1993, ApJ, 414, 474

\reference{}Shull, J. M., Harness, A., Trenti, M., \& Smith, B. D. 2012, ApJ, 747, 100

\reference{}Silk, J., \& Rees, M. J. 1998, A\&A, 331, L1

\reference{}Simpson, J. M., Smail, I., Wang, W.-H., et al. 2017, ApJ, 844, L10

\reference{}Simpson, J. M., Smail, I., Swinbank, A. M., et al. 2015, ApJ, 807, 128

\reference{}Siudek, M., Małek, K., Scodeggio, M., et al. 2017, A\&A, 597, A107

\reference{}Song, M., Finkelstein, S. L., Ashby, M. L. N., et al. 2016, ApJ, 825, 5

\reference{}Speagle, J. S., Steinhardt,C. L., Capak, P. L., \& Silverman, J. 2014, ApJS, 214, 15

\reference{}Spilker, J. S., Marrone, D. P., Aravena, M., et al. 2016, ApJ, 826, 112

\reference{}Stanley, F., Alexander, D. M., Harrison, C. M., et al. 2017, MNRAS, 472, 2221

\reference{}Stanley, F., Harrison, C. M., Alexander, D. M., et al. 2015, MNRAS, 453, 591

\reference{}Straatman, C.M.S., Labb\'e, I., Spitler, L.R. 2015, ApJ, 808, L29

\reference{}Strandet, M. L., Weiss, A., De Breuck, C., et al. 2017, ApJ, 842, L15

\reference{}Strandet, M. L., Weiss, A., Vieira, J. D., et al. 2016, ApJ, 822, 80

\reference{}Sutherland, R. S., \& Dopita, M. A. 1993, ApJS, 88, 253

\reference{}Swinbank, A. M., Harrison, C. M., Trayford, J., et al. 2017, MNRAS, 467, 3140

\reference{}Tacconi, L. J., Genzel, R., Saintonge, A., et al. 2018, ApJ, 853, 179

\reference{}Tacconi, L. J., Neri, R., Genzel, R., et al. 2013, ApJ, 768, 74

\reference{}Tadaki, K.-i., Kodama, T., Nelson, E.J., et al. 2017a, ApJ, 841, L25

\reference{}Tadaki, K.-i., Genzel, R., Kodama, T., et al. 2017b, ApJ, 834, 135

\reference{}Talia, M., Pozzi, F., Vallini, L., et al. 2018, MNRAS, in press [arXiv:1802.06083]

\reference{}Tasca, L. A. M., Le Févre, O., Hathi, N. P., et al. 2015, A\&A, 581, A54

\reference{}Teklu, A.F., Remus, R.-S., Dolag, K., et al. 2018, ApJ, 854, L28

\reference{}Thomas, D., Maraston, C., Bender, R., \& Mendes de Oliveira, C. 2005, ApJ, 621, 673

\reference{}Thompson, T.A., Quataert, E., \& Murray, N. 2005, ApJ, 630, 167

\reference{}Toft, S., Zabl, J., Richard, J., et al. 2017, Natur, 546, 510

\reference{}Tomczak, A. R., Quadri, R. F., Tran, K. H., et al. 2016, ApJ, 817, 118

\reference{}Tomczak, A. R., Quadri, R. F., Tran, K.-V. H., et al. 2014, ApJ, 783, 85

\reference{}Toomre, A. 1964, ApJ, 139, 1217

\reference{}Trujillo, I., Forster Schreiber, N.M., Rudnick, G., et al. 2006, ApJ, 650, 18

\reference{}Turner, O. J., Cirasuolo, M., Harrison, C. M., et al. 2017, MNRAS, 471, 1280

\reference{}van de Sande, J., Kriek, M., Franx, M., et al. 2013, ApJ, 771, 85

\reference{}van den Bosch, R.C.E. 2016, ApJ, 831, 134

\reference{}van der Wel, A., Franx, M., van Dokkum, P. G., et al. 2014, ApJ, 788, 28

\reference{}van der Wel, A., \& van der Marel, R.P. 2008, ApJ, 684, 260

\reference{}van Dokkum, P. G., Nelson, E. J., Franx, M., et al. 2015, ApJ, 813, 23

\reference{}van Dokkum, P.G., Bezanson, R., van der Wel, A. 2014, ApJ, 791, 45

\reference{}van Dokkum, P. G., Kriek, M., \& Franx, M. 2009, Natur, 460, 717

\reference{}van Dokkum, P.G., Franx, M., Kriek, M., et al. 2008, ApJ, 677, L5

\reference{}Veale, M., Ma, C.-P., Greene, J.E. 2017, MNRAS, 471, 1428

\reference{}Velander, M., van Uitert, E., Hoekstra, H., et al. 2014, MNRAS, 437, 2111

\reference{}Venemans, B.P., Walter, F., Decarli, R., et al. 2017a, ApJ, 851, L8

\reference{}Venemans, B.P., Walter, F., Decarli, R., et al. 2017b, ApJ, 837, 146

\reference{}Venemans, B.P., Walter, F., Zschaechner, L., et al. 2016, ApJ, 816, 37

\reference{}Vincenzo, F., Matteucci, F., Belfiore, F., \& Maiolino, R. 2016, MNRAS, 455, 4183

\reference{}Weiss, A., De Breuck, C., Marrone, D. P., et al. 2013, ApJ, 767, 88

\reference{}White, S.D.M., \& Frenk, C.S. 1991, ApJ, 379, 52

\reference{}Wisnioski, E., Forster Schreiber, N. M., Wuyts, S., et al. 2015, ApJ, 799, 209

\reference{}Wojtak, R., \& Mamon, G. A. 2013, MNRAS, 428, 2407

\reference{}Zavala, J. A., Montana, A., Hughes, D. H., et al. 2018, NatAs, 2, 56

\reference{}Zavala, J., Frenk, C. S., Bower, R., et al. 2016, MNRAS, 460, 4466

\reference{}Zhao, D. H.,Mo, H. J., Jing, Y. P., \& B\"{o}rner, G. 2003, MNRAS, 339, 12


\reference{}Zolotov, A., Dekel, A., Mandelker, N., \& Tweed, D. 2015, MNRAS, 450, 2327

\reference{}Zjupa, J., \& Springel, V. 2017, MNRAS, 466, 1625

\end{references}
\end{document}